\documentclass[sigconf]{acmart} 

\AtBeginDocument{%
  \providecommand\BibTeX{{%
    \normalfont B\kern-0.5em{\scshape i\kern-0.25em b}\kern-0.8em\TeX}}}

\copyrightyear{2021} 
\acmYear{2021} 
\setcopyright{rightsretained} 
\acmConference[ASSETS '21]{The 23rd International ACM SIGACCESS Conference on Computers and Accessibility}{October 18--22, 2021}{Virtual Event, USA}
\acmBooktitle{The 23rd International ACM SIGACCESS Conference on Computers and Accessibility (ASSETS '21), October 18--22, 2021, Virtual Event, USA}
\acmDOI{10.1145/3441852.3471218}
\acmISBN{978-1-4503-8306-6/21/10}

\usepackage{balance}
\usepackage{caption}
\usepackage{graphics}
\usepackage{xspace}
\usepackage{xcolor}
\usepackage{color, colortbl, soul}
\usepackage{tikz}

\def\S{Sec.\xspace}
\def\insitu{\textit{in situ}\xspace}
\def\ie{\textit{i.e.,}\xspace}
\def\etal{\textit{et al.}\xspace}
\def\etc{\textit{etc.}\xspace}
\def\eg{\textit{e.g.,}\xspace}

\def\aka{\textit{a.k.a.}\xspace}
\def\vs{\textit{v.s.}\xspace}
\def\First{\textit{First}\xspace}
\def\Second{\textit{Second}\xspace}

\def\Finally{\textit{Finally}\xspace}

\newcommand{\barriernum}{$12$}

\newcommand{\presec}{\vspace{-0.1in}}
\newcommand{\postsec}{\vspace{-0in}}
\newcommand{\presub}{\vspace{-0.1in}}
\newcommand{\postsub}{\vspace{-0in}}

\begin{document}

\title[Barriers and Design Opportunities to Improve Healthcare and QOL for Older Adults through Voice Assistants]{Understanding Barriers and Design Opportunities to Improve Healthcare and QOL for Older Adults through Voice Assistants}

\author{Chen Chen, Janet G. Johnson, Kemeberly Charles, Alice Lee, Ella T. Lifset, Michael Hogarth, Alison A. Moore, Emilia Farcas, Nadir Weibel}
\email{{chenchen, jgj007, kecharle}@ucsd.edu, aliceelee98@gmail.com, {etlifset, mihogarth, alm123, efarcas, weibel}@ucsd.edu}
\affiliation{%
    \institution{University of California San Diego}
    \city{La Jolla}
    \state{CA}
    \country{United States}
}

\renewcommand{\shortauthors}{Chen \etal}

\begin{abstract}
Voice-based Intelligent Virtual Assistants~(IVAs) promise to improve healthcare management and Quality of Life~(QOL) by introducing the paradigm of hands-free and eye-free interactions.
However, there has been little understanding regarding the challenges for designing such systems for older adults, especially when it comes to healthcare related tasks.
To tackle this, we consider the processes of care delivery and QOL enhancements for older adults as a collaborative task between patients and providers.
By interviewing $16$ older adults living independently or semi--independently and $5$ providers, we identified \barriernum~barriers that older adults might encounter during daily routine and while managing health.
We ultimately highlighted key design challenges and opportunities that might be introduced when integrating voice-based IVAs into the life of older adults.
Our work will benefit practitioners who study and attempt to create full-fledged IVA-powered smart devices to deliver better care and support an increased QOL for aging populations.
\end{abstract}

\ccsdesc[500]{Human-centered computing~Empirical studies in accessibility}

\keywords{Gerontechnology, Accessibility, Health -- Well-being, User Experience Design, Older Adults, Semi-Structured Interview}

\maketitle

\section{introduction}~\label{sec::intro}
Population aging is identified as a global issue in the $21$st century~\cite{un_population_aging}.
Today's information technologies have been shown to be an effective method to improve management of healthcare and life routine, leading to the enhancement of the Quality of Life (QOL)~\cite{Weiner2003, Czaja2019}.
However, accessing and using such novel technologies is often not straightforward for aging individuals, who are often therefore left behind.
While numerous systems, \eg~Electronic Health Record~(EHR) and Patient Portals~(PPs), have been widely used to ease health data management and patient-provider communications, when it comes to aging populations with multiple comorbidities, it is challenging for them to adopt, learn, and interact with such tools that can usually only be accessed through Graphical User Interfaces (GUIs) on desktops and mobile devices~\cite{Tracy2010, Heinz2013}.

Today's voice based Intelligent Virtual Assistants~(IVAs) allows users to naturally interact with digital systems hands-free and eye-free.
This makes them the next game changer for the future healthcare~\cite{Yaroslav2019}, especially among aging populations~\cite{Sergio2019}.
With the increased adoption of IVAs~\cite{alexa_marketshare, va_population}, studies have analyzed how older adults are using existing features of IVAs in smart speakers to enhance their daily routines, as well as potential barriers hindering their adoptions~\cite{Milka2020, Katherine2020, Alisha2019, Randall2018}. 
While these findings focused primarily on the older adults' experience after using existing IVA features on a specific type of smart-home device, practical needs, challenges and design strategies for integrating these devices into everyday life of older adults are still unexplored.
%
Further, while the quality of healthcare delivery and QOL enhancement is determined by both care providers and patients~\cite{ncqa_health_quality_guide}, prior work only focused on the experience on the patients' side, leaving the gulf between providers' expectations and quality of care delivery wide open.

Instead of focusing on a specific smart-home device or existing features, we take one step back and attempt to understand the design space of conversational IVAs to improve the healthcare and QOL for older adults.
We report on our semi-structured interview study with $16$ older adults (aged $68$ -- $90$), two geriatricians and three nurses from  UC San Diego Health\footnote{UC San Diego Health: \url{https://health.ucsd.edu/}}.
Unlike prior work (\eg~\cite{Milka2020}), participants have different experience levels regarding the use of IVA-powered smart devices to prevent design thinking mindset caused by past learning experience.
Through the identifications of \barriernum~barriers that older adults might encounter while managing their health and daily life, we discuss the challenges and opportunities for future research to design more accessible and usable voice based IVA-enabled technologies for aging populations.

Due to global health crisis~\cite{covid19}, we also address the impact of COVID-19 lockdown and isolation that magnifies multiple hidden barriers and needs.
We believe that our findings will benefit designers, engineers, and researchers who study and create full-fledged smart devices with built-in IVAs to deliver better health care and support an increased QOL for aging populations.

\presec~\section{Related Work}\label{sec::related}
\postsec

\subsection{Technology Adoption by Older Adults}\postsub
The Technology Acceptance Model~(TAM) suggests that acceptance and usage of technology heavily depends on the Perceived Usefulness~(PU) and Perceived Ease of Use~(PEOU)~\cite{Davis1989}.
Arning~\etal~\cite{Arning2007} studied the TAM in the context of aging and outlined how performance was different across young (focused on task efficiency), and older adults~(focused on task effectiveness).
They also found PU plays a more important role among older people compared to young adults~\cite{Arning2007}.
As for mobile technologies, Kim~\etal~\cite{Kim2016} suggested conversion readiness, self-efficacy, and peer support as three key factors affecting older adults' cognitive process. 
While learning new general technologies, Pang~\etal~\cite{Pang2021} suggested older adults appreciating self-paced learning, remote support, and flexible learning methods, instead of instruction manuals.

Despite these efforts, aging is still typically framed as a ``problem'' that can be managed by those technologies~\cite{Vines2015}. 
Existing works have focused on older adults' experience while using touchscreen-based mobile devices for well-being and self-management.
Piper~\etal~\cite{Piper2010} examined the acceptability of surface computing for health support with older adults.
They found older adults felt less intimidated, frustrated and overwhelmed when using the surface computer compared to a desktop machine, yet some gestures requiring two fingers (\eg~resize) and fine motor movement (\eg~rotate) were challenging.
The team explored different forms of tablet computers and identified these challenges especially when such devices were used as communication platforms~\cite{Piper2016}.
Doyle~\etal~\cite{Doyle2014} conducted a $5$-month deployment of YourWellness, an application that supported older adults in self-reporting and self-managing their well-being, and contributed the understanding of older adults' attitudes and behaviors regarding well-being and self-management.

Prior research also explored the adoption of wearable tracking devices by older adults.
Mercer~\etal~\cite{Mercer2016} reported how older adults considered achieving walking goals and competing with themselves as crucial self-motivators for using commercially available wearable activity trackers.
By comparing $10$ focus groups at different use stages of wearable activity trackers, Kononova~\etal~\cite{Kononova2019} found that the motivation for long-term use and maintenance were the recognition of long-term benefits of tracker use, social support, and internal motivation.
For those characterized by social isolation and loneliness, the social connectedness is considered an additional critical factor for technology adoption~\cite{Newall2019}.

Overall, this body of work focused on the barriers and needs for older adults to develop and maintain long-term use of mobile and wearable technology.
However, the majority of them are based on technologies accessed through touchscreen-based input and GUI-based output. 
In contrast, our study focuses on the older adults' use of technologies through conversational voice interface.

\presub
\subsection{Use of IVA Technologies by Older Adults}\postsub
IVAs refer to software Artificial Intelligence~(AI) agents for realizing conversational user interfaces, which can understand human speech, perform tasks and services based on input queries, as well as respond via synthesized voices~\cite{Hoy2018}.
Such assistants can be incorporated into a diverse set of mobile and wearable devices.
The embodiment of these devices can be either \textit{user attached}---usually hand-held or wrist-worn by end-users (\eg~the Siri assistant on iPhone and iWatch), or \textit{user detached}---standalone and usually affiliated to a specific environment (\eg~smart speakers and intelligent appliances with conversational capabilities).

One thread of existing research focuses on the \textit{user detached} IVAs.
A recent study~\cite{va_population} demonstrated that younger Americans aged $18$ -- $29$ are $75\%$ more likely to own a smart speaker than those over $60$. 
However, among these users, the usage frequency for older adults slightly exceeded that of young people ($46.6\%$ \vs~$43.1\%$).
Kim~\cite{Kim2021} suggested a positive overall first impression of older adults toward smart speaker based voice assistants, and indicated healthcare related questions and music streaming as the top $2$ topics that participants made in first interaction.
Trajkova~\etal~\cite{Milka2020} explored the reasons why older adults with no experience using smart speakers have difficulties finding valuable uses associated with their abilities and beliefs. 
Pradhan~\etal~\cite{Pradhan2020} additionally suggested that the practical usage of smart speakers was unexpectedly low due to reliability concerns.
Bonilla~\etal~\cite{Karen2020} conducted semi-structured interviews with older adults using Amazon Echo and Google Home, and unveiled important privacy and security concerns that caused negative reviews.
Sander~\etal~\cite{Jamie2019} designed a Wizard of Oz study and uncovered older adults' need for autonomy in terms of data management and personal health decisions.
Ziman~\etal~\cite{Randall2018} explored seniors' perceptions of Voice User Interfaces~(VUIs) and proved the feasibility of such interfaces for older adults, even though they are often regarded as opposed to adoption of new technologies.
Similarly, Pradhan~\etal~\cite{Alisha2019} explored the anthropomorphism aspects, \ie~how older adults treat a smart speakers-based IVA as a person. 
They found older adults have tendencies to anthropomorphize smart speakers when they are trying to explain and understand device behaviors, or looking for social interactions while feeling lonely.
As for smart home technology, Kolwaski~\etal~\cite{Jaroslaw2019} examined the use of smart speakers to control IoT devices by older adults, and highlighted existing cognitive and physical needs. 

Another thread of works explored the design of \textit{user attached} IVAs---which usually also support touch and speech modalities, paired with a GUI---to help older adults manage their daily routines.
Unlike user detached devices, these kinds of IVA systems are usually carried by users rather than acting as a fixed element in the environment.
For instance, Teixeira \etal~\cite{Antonio2017} and Ferreira~\etal~\cite{Ferreira2013} demonstrated the merits of speech modality while designing smartphone-based medical assistants.
Through a lab-based comparative study on text input, Schl\"{o}gl~\etal~\cite{Schlogl2013} suggested that older adults preferred to use VUI to GUI-based touch screen input.
While designing tablet based applications for supporting food intake reporting, Liu~\etal~\cite{Liu2020} suggested voice only reporting requires significant less time and is less error prone compared to the voice button reporting.
Zajicek~\cite{Zajicek2001} also suggested that the QWERTY keyboard could be illogical for older adult, making tasks such as finding and pressing keys becoming daunting.
Wulf~\etal~\cite{Wulf2014} conducted a similar study with older adults using iPhone 4s with Siri to perform basic tasks and found that participants appreciated the fast command input speed through voice. 
In comparisons, Smith~\etal~\cite{Smith2015} found that both young and older adults can input text equally fast with voice dictation, but older adults were significantly slower than younger adults when it comes to QWERTY keyboard and handwriting. 
As for word error rate, both age groups had low error rates when using physical QWERTY keyboard and voice, but older adults committed more errors with handwriting. 
Finally, Sato~\etal~\cite{Sato2011} used voice as the auxiliary output modality to augment a web interface by reading aloud text confirmation and status changes, and discussed the potentials of using voice as enhancement to the web browsing experience of older adults.

We consider our study complementary to these works, yet unique from two angles:
\First, we go beyond evaluating user experiences of existing built-in features (\eg~\cite{Milka2020}), and instead focus on the barriers that older adults might come across in their daily life and during their healthcare management. 
Ultimately, we aim to identify the challenges and opportunities for future researchers to incorporate IVAs into assistive devices based on the barriers that older adults encountered.
\Second, our results represent the combination of analyzing input from both healthcare providers and their older adults patients.
Unlike the aforementioned research that only focuses on patients, and their general use cases, providers play a key role in the healthcare delivery process, even though they might not interact directly with the IVAs and assistive devices.
We also follow Wang~\etal's finding~\cite{Wang2016} that even when older adults do not express much interests in exploring the use of mobile and wearable healthcare technologies, nor they understand the existence of these tools, they still trust and rely on their providers to know how to best manage their health.

\presub
\subsection{Integrating Voice into Healthcare Systems}\postsub
Promoting self-care and patient engagement are key features to enhance health service delivery and care quality. 
For older adults, this is magnified by the prevalence of chronic health problems and potential medical burdens that increase with age~\cite{Kim2018, Latulipe2018, Miller2016}.
PPs should have been used to promote patients' engagement with their healthcare providers and data in a variety of ways, \eg~communicating with care providers, reviewing test results, scheduling appointments, and refilling medications~\cite{Taya2015}.
But the practical use of PPs has not been widespread among older adults due to a number of barriers in their adoption~\cite{Ramirez2020, Goldzweig2013, Taya2015}. 
Niazkhani~\etal~\cite{Niazkhani2020} concluded that the barriers for managing chronic diseases using EHRs and PPs was associated with multiple factors from the patients' side (\eg~age, gender, health status, computer literacy, preference for direct communication, and patient strategy for coping with a chronic condition), the providers' side (\eg~providers' lack of interest or resistance to adopting EHRs due to overwhelming workload, lack of reimbursement, and training), the technology side (\eg~concerns of security and privacy guarantees, lack of interoperability and customized features for chronic diseases), and chronic disease characteristics (\eg~comorbidities).

To enhance the accessibility of PPs, researchers explored the use of voice as an alternative to GUI-based PPs~\cite{Takagi2018}.
Crystal~\etal~\cite{Kumah2018} outlines two characteristics of VUIs that make them a promising modality to integrate within the healthcare delivery infrastructures.
\First, voice represents a natural choice of medium for eye-free and hands-free interaction between aging individuals and digital information, especially those with visual and mobility impairments~\cite{Jennifer2011}.
\Second, studies have shown that conversational speech input is up to $3\times$ faster compared to keyboard for English, while the error rate is $20.4\%$ lower~\cite{Ruan2016}. 
The increased efficiency compared to keyboard input is particularly significant among older adults who may have less familiarity with computing technologies, or might experience accessibility issues due to declining physical abilities.

With the benefits of traditional dictation-based VUIs, conversational VUIs powered by IVAs allow users to naturally tell computer systems what to do without the rigid requirement of syntax-specific commands~\cite{Porcheron2018}.
A full review of conversational agents in healthcare is beyond our scope, however, we recommend readers to render Laranjo \etal's work~\cite{Laranjo2018} as the starting point.
The integration of IVAs into the healthcare system promises to form an ``alliance'' and  create the ``rapport'' with patients through natural conversation that is expected to be beneficial to treatment outcomes~\cite{Burebe2020}.
Prior works have explored the use of IVAs to automate and promote the health data interactions on the patients' side.
A well-known example is the Interactive Voice Response System~(IVRS), an automatic telephone-based system that interacts with patient callers for the purpose of triaging~\cite{James1997}.
Due to constraints on the information exchange and the depersonalization of customer service, IVRS has generally been viewed unfavorably~\cite{Kumah2018}.
While several health systems are exploring ways to bring voice based IVAs into care delivery processes, most existing features are essentially reactive ``question-and-answering'' model, leading to poor user experience, and making applications, \eg~\emph{Blood Pressure Logs}~\cite{blood_pressure_logger}, only being used as dictation data storage.
This is impractical, especially for those having memory impairment and might forget the time for measuring vital signs without proactive reminders. 

\presec
\section{Methods}~\label{sec::methods}\postsec
We employ a {\it user-centered design}~\cite{Elizabeth2001, Stephanie2006} to better understand the interconnections across patients, providers, the underlying healthcare system and its existing tools (\eg~PPs).
Our study was approved by the Institutional Review Board~(IRB).

\presub
\subsection{Participants}~\label{sec::methods::participants}\postsub
Geriatric care delivery and the subsequent enhancement of QOL is a collaborative task between patients and providers.
Therefore, we designed semi-structured remote and in-person interviews~\cite{Lisa2008} with older adults (age, $\mu = 76.56$, $\sigma = 6.97$) and healthcare professionals, respectively (see Appendix~\ref{appenx:demo} for participants' demographic data). 
Participants were recruited through UC San Diego Health and rewarded with a \$$20$ gift card after the interview.
Due to the nature of needs finding research that is different to user experience investigations, we include participants who claimed themselves as ``not used but knew about IVA''~($3$~participants, $19$\%)  and ``neither used nor knew about IVA''~($4$~participants, $25$\%). 
This helps us minimize participants' design thinking mindset caused by past user experience. Our study only includes patients living independently or semi-dependently.

\presub
\subsection{Remote Semi-Structured Interviews}~\label{sec::methods::interviews}\postsub
In response to COVID-19 pandemic and increased risks of contagion with older adults~\cite{covid19}, we conducted interviews with older adults remotely over the phone. 
Since our interviews were semi-structured, we did not strictly follow a formalized list of questions. 
Instead, we used open-ended guiding questions to encourage participants to tell us about their stories and experiences.
Participants were not expected to answer the questions explicitly. 
Our team, composed of human-computer interaction, pervasive healthcare, and geriatrics experts, collaboratively designed a list of topics and kept it consistent across all participants (see Table~\ref{tab::protocol}). 
Rather than explicitly summarizing their arguments, participants tended to bring up arguments using personal experiences and specific stories.
Notably, for those who were not familiar with IVAs or without previous experience of using an IVA, we first introduced the conversational VUI and IVA conceptually.

\begin{table}[t]
    \centering
    \includegraphics[width=0.5\textwidth]{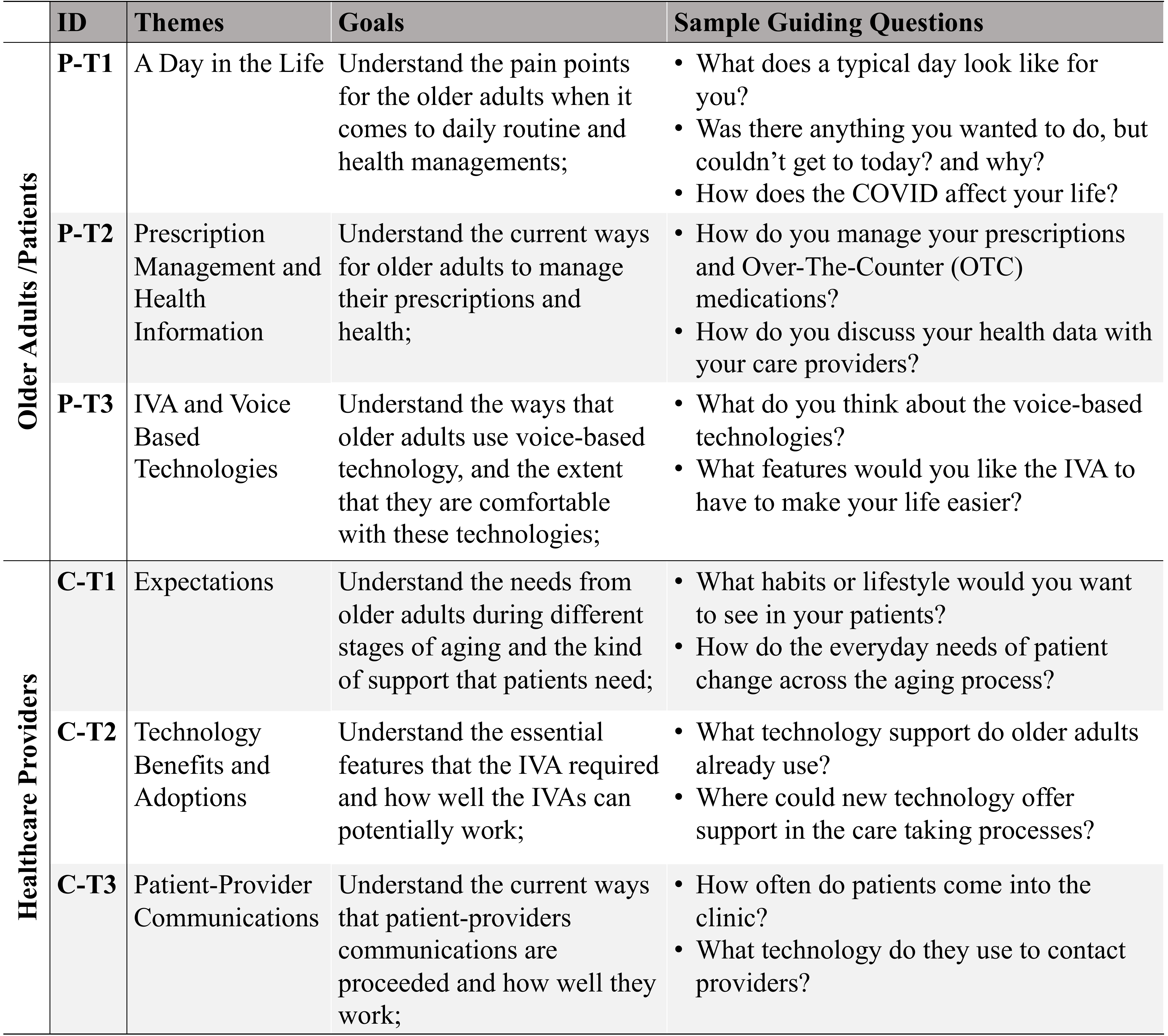}
    \Description{Themes and guiding questions collaboratively designed with geriatric professionals. We use P-T#and C-T#to indicate Patients' themes and Clinicians' themes respectively, \eg~P-T1 is the first theme discussed with older adults.}
    \captionof{table}{Themes and guiding questions collaboratively designed with geriatric professionals. We use \textbf{P-T$\#$} and \textbf{C-T$\#$} to indicate \textbf{P}atients' themes and \textbf{C}linicians' themes respectively, \eg~\textbf{P-T1} is the first theme discussed with older adults.}
    \label{tab::protocol}
    \vspace{-1.0cm}
\end{table}

\presub
\subsection{Data Analysis}\label{sec::methods::analysis}\postsub
We transcribed the recorded audio clips and manually edited/corrected the transcriptions as needed. 
We manually coded the qualitative data using inductive and deductive coding approaches~\cite{Satu2008} rooted in grounded theory~\cite{strauss1994grounded}.
We employed Optimal Workshop\footnote{Optimal Workshop: \url{https://www.optimalworkshop.com}} to facilitate the analysis by  efficiently tagging participants' responses and generate themes with corresponding participants' responses.
Upon conflicts, researchers engaged in multiple discussions to iteratively compare and reconcile the codes, until all researchers were satisfied with the outcome~\cite{Nora2019}.
We finally outlined individual user stories to characterize the particular themes that emerged from our analysis. 
Specific user stories helped us to better understand the point of view of participants, and thus synthesize their barriers and needs.

\presec
\section{Findings: Barriers to Manage Healthcare and Quality of Life}~\label{sec::results}\postsec
Table~\ref{tab::findings} shows our findings from patients' and providers' perspective.
In this section, we will reason barriers by connecting to participants' real stories and experience.
\begin{table}[t]
    \centering
    \includegraphics[width=0.5\textwidth ]{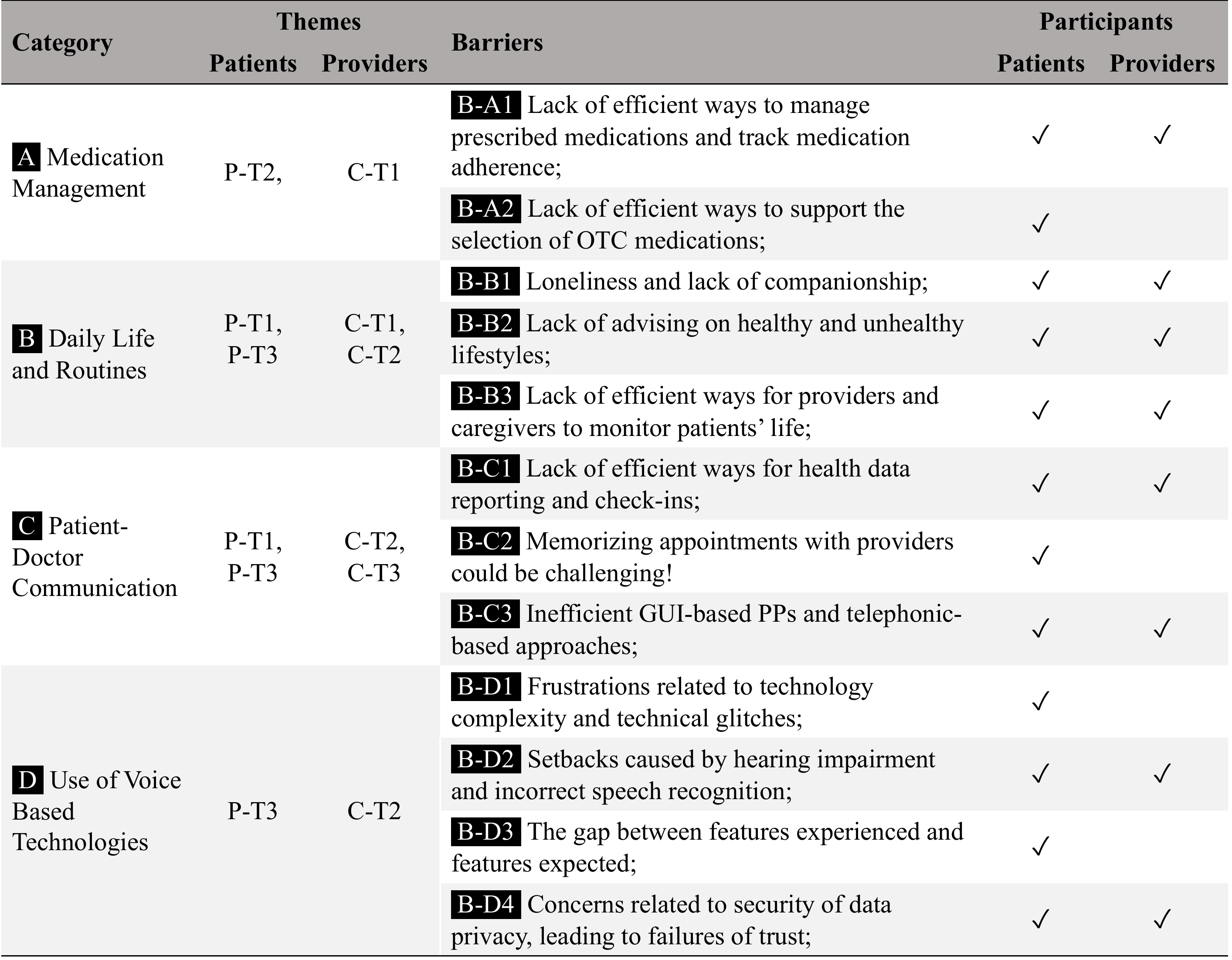}
    \Description{Overview of barriers emerging from patients and providers across four categories. The indices of the theme scan be seen in Table 1.}
    \captionof{table}{Overview of barriers emerging from patients and providers across four categories. The indices of the themes can be seen in Table~\ref{tab::protocol}.}
    \label{tab::findings}
    \vspace{-1.0cm}
\end{table}
\presub
\subsection{Medication Management}~\label{sec::finding::medication_management}\postsub
\noindent{\bf [B-A1]~\emph{Lack of Efficient Ways to Manage Prescribed Medications and Track Medication Adherence}}

\noindent 
Both providers and patients described how older adults often forget to take their medications, misunderstand dosage, or underestimate the importance of it.
Their declining cognitive health also makes it increasingly challenging to keep track of their medications as aging~[C2].
To address these, the After Visit Summary~(AVS) through PPs, \eg~MyChart\footnote{Epic MyChart is used by UC San Diego Health: \url{https://www.mychart.com}.}, is widely used to deliver detailed information about the medication regimen. 
Both geriatricians found this to be helpful, especially for \textit{``those with memory issues or lack of understanding''}~[C3].
While Federman~\etal~\cite{Federman2018} unveiled the inflexibility and rigidity of today's AVS from the clinical leaders' perspectives, we discovered how inaccessibility of AVS might affect older adults' ability to manage medications from $3$ different perspectives:

\textbf{(1)} Despite the intensive detail on AVS, older adults mentioned their confusion about the potential side effects of the prescribed medications, the appropriate dose, duplicated drugs, or drug interactions.
P13 complained: {\it ``I'm gonna talk to [the doctor] about cutting down [the medications] because I think some of them are duplicated. Some [medications] do the same thing or they have a crossover effect with one required for this ailment. And also, I think several medications prescribed to me could be used for the same issue''}.
As for the importance of understanding the potential side effects, P13 reported an unpleasant experience: {\it ``... When I took Gabapentin, [my provider] had me taking such a large dose. After taking it, I told them I couldn't take it anymore because I was having suicidal ideas and stuff like that!''}~\footnote{P13 realized that this happened, because of the disappearing of suicidal thoughts after stopping taking Gabapentin.}

\textbf{(2)} 
Due to the complicated regimen with multiple pills taken at different time of day [C3] and the effectiveness of certain  medications being highly dependent on the time taken (\eg~pills for thyroid issues are not as effective unless they are taken half an hour before a meal~[C1]), older adults mentioned the difficulties on planning out their medication schedules.
For example, P5 complained: {\it ``I was supposed to take [a pill] $4$ times today and dissolve it in water. But I have to have an empty stomach, and I can't take it within $3$ hours of taking other pills or eating...[it was] a nightmare!''}

\begin{figure}[t]
    \centering
    \includegraphics[width=0.5\textwidth]{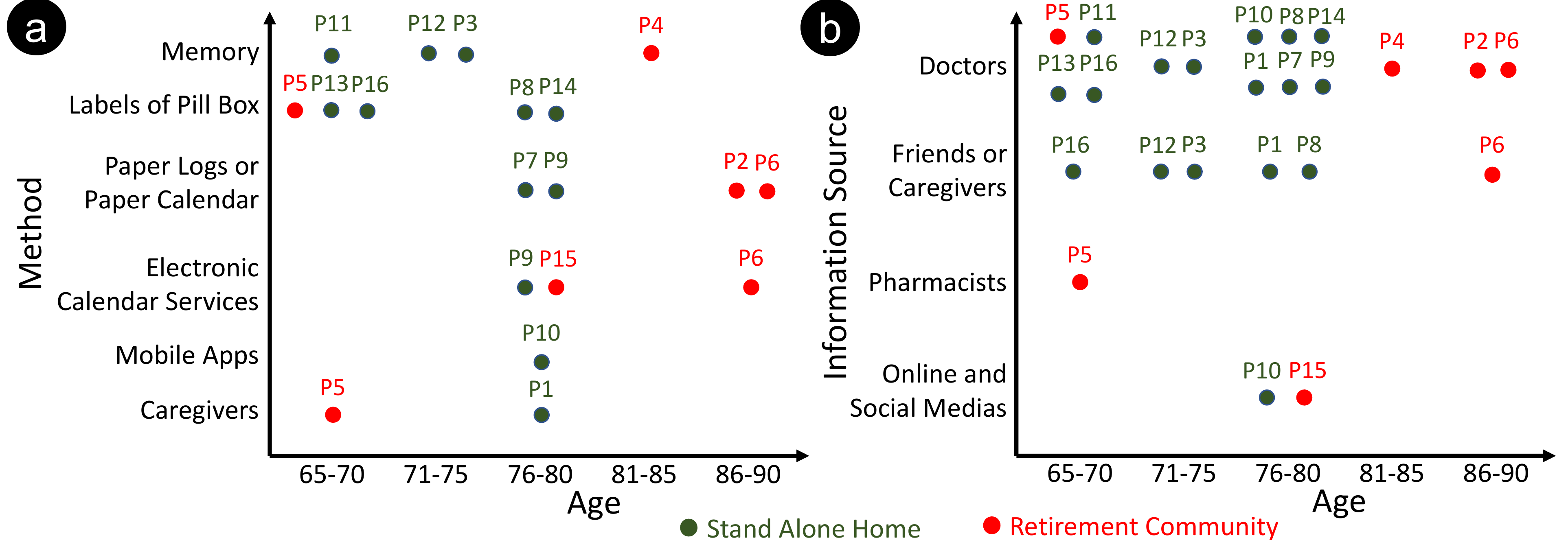}
    \vspace{-0.8cm}
    \Description{Figure 1a illustrates the scatter plots of 6 different methods that older adults used to remember their medication schedule. For 65 to 70 years old participants, P11 used memory P5, P13, and P16 used labels of the pill box, P5 relied on caregivers. For 71 to 75 year old participants, P12 and P3 used memory. For 76 to 80 years old participants, P8 and P14 used the labels of pill box, P7 and P9 used paper logs or paper calendar, P9 and P15 used electronic calendar services, P10 used mobile apps, P1 relied on caregivers. For 81 to 85 years old participants, P4 used memory. For 86 to 90 years old participants, P2 and P6 used paper logs and paper calendar, and P6 used electronic calendar services. Figure 1b illustrates the scatter plot of sources of information on OTC medications or supplements. For 65 to 70 years old participants, P5, P11, P13, P16 used relied on doctors; P16 relied on friends or caregivers; P5 relied on pharmacists. For 71 to 75 years old participants, P12 and P3 relied on doctors; They also relied on friends or caregivers. For 76 to 80 years old participants, P10, P8, P14, P1, P7, P9 relied on doctors; P1 and P8 relied on friends or caregivers; P10 and P15 relies on online and social medias. For 81 to 85 years old participants, P4 relied on doctors. For 86 to 90 participants, P2 and P6 relied on doctors; P6 relied friends or caregivers.}
    \caption{(a)~$6$ different methods that older adults used to remember their medications schedule; (b)~Sources of information on Over-The-Counter (OTC) medications or supplements. Note that there are cases when a particular patient uses multiple ways to remember the medication schedule or acquire OTC medication information.}
    \label{fig:medication}
    \vspace{-0.8cm}
\end{figure}

\textbf{(3)} Both patients and providers mentioned that today's technologies do not provide effective features for older adults to memorize and track their medications, forcing them to use alternative methods to address the problems.
Figure~\ref{fig:medication}a illustrates how the $16$ older adults in our study used $6$ different methods to memorize the medication schedule.
Four participants were confident in their ability to keep track of medication regimen solely based on memory as they believed \textit{``things have gotten into a pattern over the years"}~[P8].
However, this method is challenging in practice due to lapses in memory that come with everyday living.
P3 stressed the occurrence of such situations because of insufficient sleep.
Four older adults used handwritten notes to remind themselves. 
However, organizing and memorizing the locations of these fragmented notes proved to be difficult, mainly due to the age-associated memory impairment~\cite{Marighetto2001, Park1998}.  
For example, P6 used \textit{``little notes here and there, but many times [she] couldn't find [them],''} and P5 who has over $10$ medications often forgets to write down which ones were already taken \textit{``because [she] get[s] distracted doing something else.''}
Pill boxes were another method of managing medications used among older adults.
While C4 talked about solutions with sophisticated designs that release pills based on a timer, these tend to be expensive and, thus, less accessible.
Three participants used software applications (\eg~web-based calendar services and smartphone applications) to remind themselves. 
However, interaction rigidity occasionally raised frustration. 
For example, P11 complained: {\it ``If you don't acknowledge the reminder, you may not get another one until you know...you need to go back and acknowledge the reminders. So, if I take a medication, but don't acknowledge the current reminder. The app might not remind me about the evening one.''}\enspace
Two participants relied on their caregivers, who had full responsibility of their medications schedule. 
However, caregivers themselves might find it difficult to stay organized and be on top of the older adult's regimen, as they typically juggle other responsibilities (\eg~adult children with their own families) or their own medications as well (\eg~aging spouses). 

Unfortunately, upon failing to manage prescribed medications, some older adults reported to simply start to take them whenever they remembered, instead of consulting their providers. This behavior is risky and can potentially lead to considerable morbidity, mortality, and avoidable healthcare costs~\cite{Bosworth2011}.
\text{P13} explained one action that he would typically take after forgetting the medications: {\it``It depends on what the medicine is that I skipped. Usually I just take the next dose. It depends on how far I am from the time I was supposed to take it. So if I'm within several hours, then I'll go back and take it. But if it's more time, then I'll take it when the next dose is due.''}

\vspace{1em}
\noindent{\bf [B-A2]~\emph{Lack of Efficient Ways to Support the Selection of OTC Medications}}

\noindent All older adults have been routinely taking Over-The-Counter~(OTC) medications or supplements.
Similarly as with their reminding strategies, their decision of taking OTC medications was based on the information they received from different sources~(see Figure~\ref{fig:medication}b). 
However, the information sometimes could be incorrect, especially when older adults select OTC medications based on non-professional advice (\eg~online and social media).

All patients took suggestions from their doctors or people in their social circle with experience working in the healthcare industry, \eg~P8 who mentioned: \textit{"I have a friend who is an ex RN [Registered Nurse]. And she suggested that I should take [these supplements], so I am taking [them] - It's supposed to boost your immune system"}. 
However, we found that two participants made decisions based on suggestions from discussion groups on social media. 
Although taking OTC supplements based on non-professional advice can be potentially detrimental to one's health, and there is sometimes little scientific support on the benefits promised by supplements vendors~\cite{noauthorOTC2014, agingCare2020}, we found that all participants in our interview underestimated the potential side effects and the importance to always engage healthcare professionals in their decisions.

\presub
\subsection{Daily Life and Routines}~\label{sec::finding::routines}\postsub
\noindent{\bf[B-B1]~\emph{Loneliness and Lack of Companionship}}

\noindent Geriatricians identified how, {\it``loneliness is common in older people''}~[C1].
This echoes Demakakos~\etal's finding that loneliness progresses non-linearly across middle and older age and peaks among persons that are over 80 years old~\cite{Demakakos2006}.
In fact, providers reported on how they often receive calls from older adults {\it ``not because they have some medical issues, but they want someone to talk [\eg~recent trips and hobbies]''}~[C3].
This also echoes P13's comments: {\it``[During the pandemic], I'll just select someone and call them and say thank you. And a lot of times saying thank you turns into a conversation about why I do it and who I do it to.''}

Due to the recent impact of COVID-19 and the repeated lockdown situations, older adults, especially those living in a retirement community, stressed these problems.
With higher risk for older adults, policies (\eg~mask wearing and social distancing practices) are usually strictly enforced in retirement communities~\cite{retirement_community_covid19}.
Some older adults complained: {\it``I felt like I was in a prison ... that my freedom is limited''} [P15] and were worried about their community peers: {\it``[they] cannot go out moving, socializing''}~[P5] or {\it``visiting one another in the apartment [is not possible]''}~[P2]. 
For instance, if older adults wanted to take a walk occasionally, they needed to be alone, and they had to worry about falling and not getting help right away, since no one was accompanying them~[P2 and P13 (on behalf of his wife)]. 
Another example was P10, who had to start prescribed anti-depression medications to overcome the consequences of lockdown isolation.

While it is clearly an issue for older adults living in a retirement community, and less so for those living in a stand-alone home, older adults (\eg~P8) generally missed the daily interactions with friends and peers from social clubs. 
For example, P9 expressed the desire for in-person social interactions: {\it``I would love to go to [in-person] meetings instead of doing Zoom meetings. [Before COVID-19] I could have visited [my friends] and had lunch with them. We could have hung out in the evening, that kind of thing. But sadly we're not doing that now.''}

Social isolation is not the only reason that social interactions are limited.
Mobility impairment due to a number of health conditions could also result in similar barriers.
For example, after a $1$-month hospitalization experience, P13 complained that: {\it ``[the hospitalization] impeded my ability to meet with people the way I want to because I can't get around the way I once did.''} 

\vspace{1em}
\noindent{\bf [B-B2]~\emph{Lack of Advising on Healthy and Unhealthy Lifestyles}}

\noindent Providers reported on the importance to take into account patients' lifestyles when delivering care, but they also commented on how older adults usually lack awareness regarding {\it what lifestyle should be avoided}.
From their perspective, targeted interventions would help patients to foster a good lifestyle. For example, C3 noted how some patients are less aware of the needs for physical exercise: {\it``A lot of times we provide them with educational materials [with] recommendations like exercising $30$ minutes a day for $3$ times a week. Here are some suggested exercises: [... some example exercises ...]. And we check up on them like 'hey did you exercise like I told you to?'''}
C5 also stressed that some patients who should drink more fluids sometimes forget: {\it``We need to remind [those who are prone to having urinary incontinence] to drink more fluids.''}\enspace
Nurse C3 highlighted the importance by indicating the potential consequence: {\it ``sometime, like congestive heart failure, you don’t want them to drink
[alcohol], but in general a lot of what I noticed is that there is a trend that it is either [older adults] forget and drink, or, they just don’t think about it […] So a lot of ER visits happen because of dehydration.``} 

Similar to medication management, providers preferred to put this supporting information for healthy behavior in the AVS, where patients could access it through the PPs along with additional educational materials [C3]. 
However these turn out to be less effective, possibly due to the ambiguity of AVS and the lack of \insitu reminding from their providers.
This is consistent with the accessibility barriers discussed above (see \S~\ref{sec::finding::medication_management}) and identified in \cite{Federman2018}. 

\vspace{1em}
\noindent{\bf [B-B3]~\emph{Lack of Efficient Ways for Providers and Caregivers to Monitor Patients' Life}}

\noindent Both providers and patients found the lack of efficient ways for providers to monitor the older adults' life and activities frustrating, and recognised how sometimes this can lead to life-threatening accidents.
For example, C2 expressed the concerns of lack of efficient ways to detect accidental falls in real-time: {\it`` if [my patient] is falling in between the night, I would like to know [immediately].''}\enspace
P14 mentioned how due to forgetting to take blood pressure medications, he once risked a life-threatening accidents:
{\it``I realized I forgot to take my medication before I go to bed. [On the second day], after taking my blood pressure, it was very high. So I was scared, went upstairs, woke up my wife and my son to bring me to [the emergency room].''}\enspace
If caregivers are available and around, it is often possible to get the timely emergency service that is needed. However, for those living alone and without access to caregivers, frustration and dangerous situations can potentially arise.

Providers also complained of such barriers when it comes to those with mental health conditions~(\eg~delirium, which commonly occurred among those in hospital settings~[C1]).
One common measure in those cases was to assign nurses or instruct caregivers to take care of them. However, some older adults, especially those transitioning from independent to dependent care, can be very stubborn and not willing to receive any help~[C4].

\presub
\subsection{Patient-Provider Communication}~\label{sec::barrier::communication}\postsub
\noindent{\bf [B-C1]~\emph{Lack of Efficient Ways for Health Data Reporting and Check-Ins}}

\noindent Allowing patients to periodically report health data, and providers to actively check-in on older adults' conditions emerged as two critical aspects in patient-provider communications, especially for chronic disease prevention, detection, and treatment.
For example, nurse C5 mentioned how periodically supporting patients to self monitor blood pressure and glucose level would be extremely helpful for the treatment of diabetes.

To achieve independent health data monitoring for older adults, providers prefer to use PPs and encourage their patients to self-report their health data measurements through secure messaging.
However, this is error-prone and older adults find it difficult and highly inefficient.
C5 explained: {\it ``[the patients] are asked to either take a $7$ -- $10$ days' amount of values, [and then use] MyChart to message [the provider] or call [the provider] with the values.''}\enspace
Instead, some patients simply prepared a notebook for logging their health data to then bring to their provider, yet sometimes they would forget to do so. 
For example, P13 explained: {\it ``I have a little book, a little calendar that's part of it. So I put the systolic number over [diastolic] in there. And I just write them down everyday...[But] I had a couple of times when I forgot to write...''}\enspace
Using these asynchronous methods introduce delays in the reporting of abnormal measures to providers, potentially leading to negative impact on timely treatment. 
Echoing this barrier, P13 and P5 explicitly brought up medication reminders and voice-based approaches for supporting health data logging. 
P5 explained that: {\it``I have to weigh myself everyday and take my blood pressure and my sugar, and keep all that down. And my friend said that it would be a great thing for me to be able to have Alexa keep that information and then just be able to transfer to MyChart, instead of me having that typed all in for the doctors.``}

Providers also believed that mental and social health check-ins would be useful to reduce potential mental impairment caused by lack of companionship, but this responsibility is usually offloaded to the caregivers' side, which is inefficient and impractical for those without caregivers. 
For example, C3 mentioned that {\it``sometimes you'll have family members who just check in once a week to say like, how are you doing?''}\enspace
While this seems insufficient, increasing the check-in frequency can be difficult due to the lack of time and energy of those caregivers, similar to asking caregivers to manage older adults' medications (see \S\ref{sec::finding::medication_management}).

\vspace{.5em}
\noindent{\bf [B-C2]~\emph{Memorizing Appointments with Providers could be Challenging}}

\noindent Providers assumed that older adults were able to use PPs for appointments and calendar reminders.
However, appointments are routinely being forgotten by older adults.
P5 reported having trouble remembering the date, and brought up one example: {\it``I'm always forgetting when my appointments are. Once I showed up for one on Monday, but it was supposed to be on Tuesday.''}\enspace
To cope with this issue, some older adults relied on providers' reminders. 
P1 explained that {\it``the doctors would take care of all that for you. They call you the night before. Once confirmed, they send you questionnaires on MyChart. [You'll] get it all filled out when the doctor sees you...''}\enspace
However, for those whose providers do not offer such services, memorizing appointments can be somewhat challenging. 
Also, we already showed how access to PPs such as MyChart is often problematic and, therefore, often not the right solution for older adults.

\vspace{.5em}
\noindent{\bf [B-C3]~\emph{Inefficient GUI-based Patient Portals and Telephone-Based Approaches}}

\noindent Some participants found it challenging and inefficient to use GUI-based PPs and telephone-based approaches, which are two main approaches used by clinicians for asynchronous and synchronous communications.
For example, P11 felt that typing and interacting with MyChart is cumbersome, especially when it comes to mobile devices. 
P5 also complained about the complexity of setting up video meetings: {\it``Another thing that I cannot figure out is how to do a video conference with the doctor. I keep trying. And the thing will say you have not downloaded this. Why I have not downloaded it on my tablet and my phone?! So I get extremely frustrated with the system...''}~

On the providers' side, C3 brought up several general frustration that older adults encountered: {\it ``We do have patients who are like, I don't know how to use a computer; like, I don't know this MyChart; I lost the password; I don't know how to send something. [...] Sometimes it's trying to walk them through the technological aspect of how to get their care or simply reverting back to less technologically advanced [systems], like using the phone to communicate or using a letter through mail.''}

This introduces an important tension between ease-of-use on the patients' side, and availability of features on the providers' side. This is particularly magnified when it results in the older adults not being able to accomplish the goals defined by the particular feature (\eg~a tele-health visit). 
Provider C5 confirmed this issue: {\it``most of the patients [who are] not actively using MyChart ... it's their adult children that are actually interacting with us on their MyChart.''}

To overcome this situation, some participants preferred to simply call the providers over the phone to ask for medication refill requests.
Phone-based communication was also used by our participants for other care-related queries, and in general the triage nurse would make a final decision in terms of the patient needing to talk to a doctor or not, based on their internal protocol. 
However, as noted by C1, such methods sometimes are limited: {\it ``The level of service we have here is that they have a triage nurse. That says [based on a predefined protocol], do the patients need to go to urgent care? do they need to go to the ER? or do they need someone to call them back? [However] in almost no cases that the doctor's gonna call back.''}

\presub
\subsection{Use of Voice Enabled Technologies}~\label{sec::barrier::reliability}\postsub
\noindent{\bf[B-D1]~\emph{Frustration related to Technology Complexity and Technical Glitches}}

\noindent Frustration arises when older adults encounter technical glitches, which leads to decreased engagement.
For example, P6 emphasized the importance of guidance, feedback and ease-of-use: {\it``some technology is touchy and it's also sometimes complicated for old people. So it has to be easy, and not break down. When it doesn't work it's very frustrating and people like me don't know what to do. I'm not that tech savvy.''}\enspace
P11, who has experience using Amazon Echo, pointed out that a failure message reporting on {\it how to solve the problem} will be more helpful than {\it what and where the problem is}.
For example, the rigid error message, \eg~{\it ``there was a problem with the requested skill's response''} was not helpful and only resulted in additional frustration.

Upon failures, some older adults seek help from their friends or caregivers.
For example, P1 simply {\it``run to [his son, when encountering technical difficulties] and having [his son] explained what [he] just did or didn't do.''}\enspace
This implies that without an inner circle of people (\eg~caregivers and family members) who know technology and are proficient with computers, older adults may eventually abandon the device.

\vspace{.5em}
\noindent{\bf[B-D2]~\emph{Setbacks Caused by Hearing Impairment and Incorrect Speech Recognition}}

\noindent Providers also mentioned the potential barriers for older adults caused by hearing impairment, specifically when it comes to patient-doctor communications (\textbf{C-T3}).
Nurse C3 worried that such situation might cause patients to abandon using general voice enabled technologies, as similar situations occurred while talking on the phone:
{\it ``I have a patient who is hard of hearing. He is the one who calls me about once every week to discuss the same medications over and over again. Because he's hard of hearing...often times he's yelling into the phone like...everyone knows when I'm talking to him, because they're like...oh, she's yelling. So that would be one of the barriers.''}~

When it comes to the use of voice to interact with devices~(\textbf{P-T3}), two participants pointed out the frustration caused by the lack of understanding of the user's speech. 
P11 explained his wife's user experience: {\it``She asks Alexa, `please play Willie King', a blues singer. It would come back and say, `here's music by Willie Kay.' It was not correctly parsing what she wanted, and that's very frustrating, making you not want to use it anymore.''}
Although Ma~\etal~\cite{Ma2017} suggested the word misdetection rate is similar between young and older adult ($7.16\%$ \vs $4.57\%$), the misinterpretations of key phrases might introduce input failures. 
Moreover, misinterpretations might be worse for those with significant age-related cognitive decline, which might be caused by various vocal characteristics, \eg~pauses, hesitations~\cite{Kobayashi2019}.

\vspace{.5em}
\noindent{\bf[B-D3] \emph{Gap between Features Used and Additional Features Needed}}

\begin{figure}[t]
    \centering
    \includegraphics[width=0.5\textwidth, page=1]{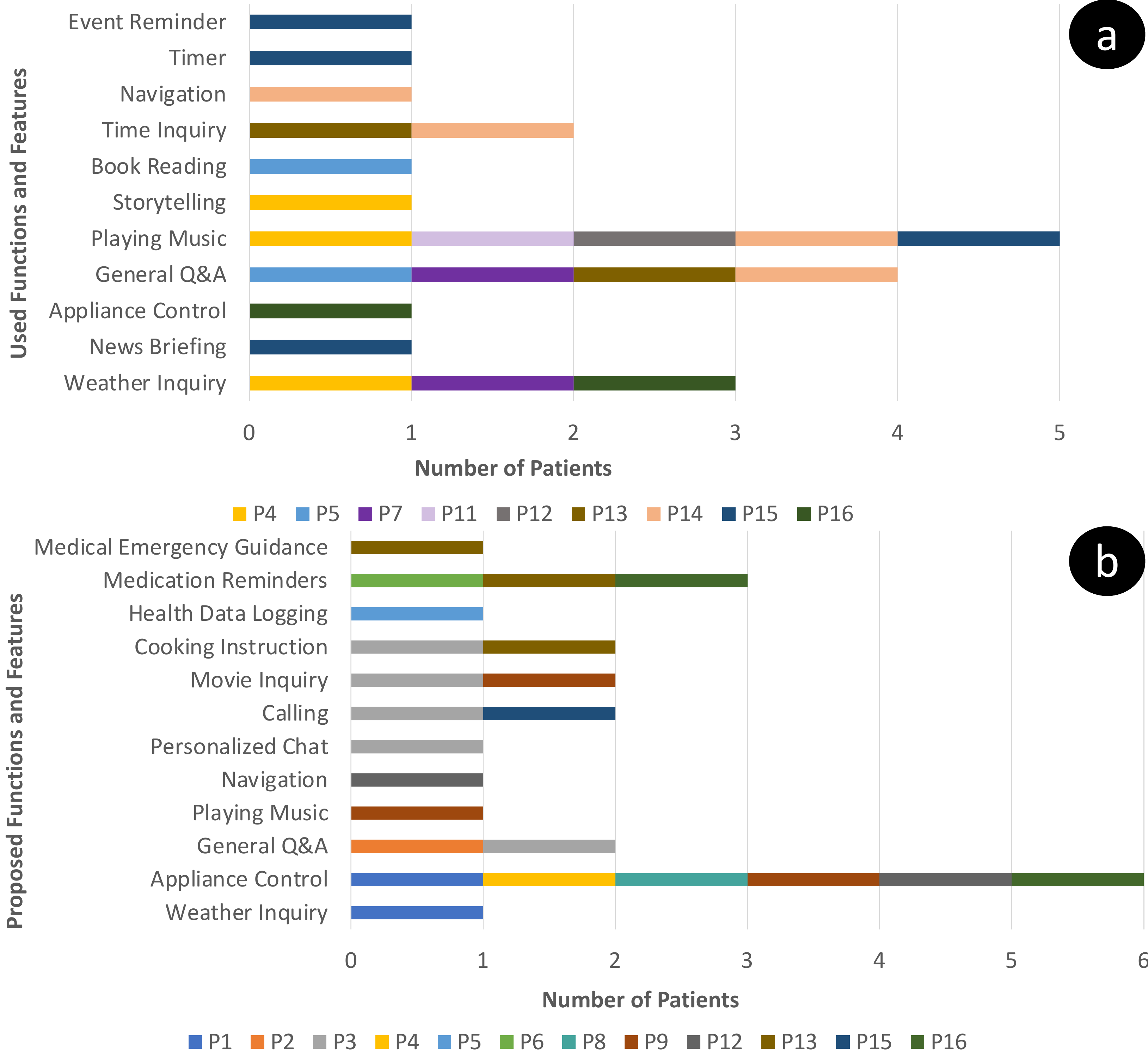}
    \vspace{-0.8cm}
    \Description{Figure 2a illustrates the technology features that older adults have used. Event reminder: P15; Timer: P15; Navigation: P14; Time inquiry: P13 and P14; Book reading: P5; Storytelling: P4; Playing music: P4, P11, P12, P14 and P15; General Q\&A: P5, P7, P13, P14; Appliance control: P16; News briefing: P15; Weather inquiry: P4, P7, P16 Figure 2b illustrates the technology features that older adult expected to have. Medication emergency guidance: P13; Medication reminder: P6, P13, P16; Health data logging: P5; Cooking instruction: P3, P13; Movie inquiry: P3, P9; Calling P3, P15; Personalized chat: P3; Navigation: P12; Playing music: P9; General Q\&A P2, P3; Appliance control: P1, P4, P8, P13, P12, P16; Weather inquiry: P1.}
    \caption{Technology features that older adults \underline{have} used (a) and \underline{would like} to use (b) in conversational voice assistants. We only include participants self-reporting ``had experience using [IVA]'' in (a). Four older adults did not propose constructive design ideas under this theme and, thus, are not included in (b).}
    \label{fig::functions_features}
    \vspace{-0.6cm}
\end{figure}

\noindent
While discussing \textbf{P-T3}, we attempted to understand features that older adults have already tried or would like to try~(see Figure~\ref{fig::functions_features}).
Surprisingly, despite the existence of voice applications designed to manage patients' medications (\eg~\cite{pill_me}), none of the older adults had tried using IVAs directly for their healthcare purposes.
Most features that older adults wished to have are already provided to some degree by mainstream IVA-powered devices. 
We infer that the gap between the features used and the new features needed could be possibly caused by the highly fragmented nature of skills, the difficulty of finding new ones that meet the user's expectations, the process of setting up features, and the fact that they are not integrated into existing patients' EHRs and PPs.

\vspace{.5em}
\noindent{\bf [B-D4]~\emph{Concerns Related to Security and data Privacy, leading to Failures of Trust}}

\noindent When discussing the concerns of voice based IVAs, three patients and two providers mentioned worries related to data security and transparency in terms of how voice data would be used.
Although the majority of IVA vendors have designed countermeasures to ensure security and privacy, participants were not convinced that their private speech data would not be leaked illegally.
P8, who worked in the legal field before retiring, brought up her concern regarding how it is {\it ``unclear  who's getting the data of what's said and what they're using it for.''}\enspace
P1 shared similar feelings, and lost faith on IVAs after hearing his friends' discussions: {\it``[My friends] say that Alexa is extremely dangerous. All [my private speech] is recorded outside. People can listen to what's going on in your home illegally, and I guess they have QAnon doing it.''} 

Providers also raised similar concerns: {\it``[using smart speakers] worries me sometimes [it feels like] I'm constantly being monitored. I know that's not what the intention is, but on the back of my mind, that's just something that happens.''}~[C3]
Such sentiments echo Bonilla~\etal's finding~\cite{Karen2020}, and indicate the need for more understandable privacy assurances than what is currently published by vendors and skill developers.
\presec
\section{Discussion}\label{sec::discussion}\postsec
In this section, we discuss the design challenges and opportunities by linking our findings in \S\ref{sec::results} with state-of-art voice based IVA technologies.

\presub
\subsection{Design for Interactions: Overcoming Ambiguous Information Input and Output}~\label{sec::discussion::interaction}\postsub
A successful interaction requires technology to understand the user's queries, and vice versa.
Ambiguous information output occurs when users cannot obtain the expected information, causing failures to perform the subsequent task based on the system's output. 
On the other side, misinterpretation of the information input happens when the system cannot understand the user's intents.
While these are well known problems for VUIs~\cite{Sergio2019}, when considering aging individuals these issues become critical.

Our data from \S~\ref{sec::barrier::reliability} outlines how older adults are frustrated when using IVAs (\textbf{B-D1}) and how this causes setbacks related to some of their physical impairments (\textbf{B-D2}). 

Specifically, we found that a major reason that caused lots of frustrations for both providers and patients with past IVA use during normal operation is the failure of speech recognition ~(\textbf{B-D2}), even when users are native English speakers and do not have significant impairment related to audio actuation systems.
Chandel~\etal~\cite{Chandel2013} speculate that completely addressing this issue is challenging, and that even a robust speech recognition platform will never achieve $100$\% reliability. 
In the healthcare setting this is even more challenging, given that understanding health and medical-related questions is more difficult, yet critical for aging populations~\cite{Mrini2021}.

As for information output, our data show that participants are generally happy during normal operation, yet are disappointed when it comes to setting up the device and when they have to deal with failure management.
P6's noted how {\it ``[the system should be] set up by somebody else, and then it should just work.''}~
As shown in \S~\ref{sec::barrier::reliability} (\textbf{B-D1}) a key reason for these failures is the rigidity of the audio output. 
It only signals the occurrence of the technology problem, and does not indicate participants \textit{how} to troubleshoot the errors.
Participants in our study also reported how the process of setting up the device is an ``impossible'' task.
We believe that this is due to the complex setup process, which usually involves the use of additional technology (\eg~to setup an Echo speaker and complete the authentication and network setup steps, users need to download the Alexa App on their mobile phone). The rigid audio output command discussed above also contributes to the problem, causing some steps during this setup procedure to often become ambiguous.
Older adults in our study reported how solving these problem often required seeking help from their family members. For those living alone, failing the initial set up phase might cause them to completely abandon the technology.

One solution that can alleviate ambiguous information and misinterpretations is to design \emph{voice-first} interfaces (and not \emph{voice-only} interfaces), where alternative input-output modalities, \eg~touch screen, is also available to receive more explicit input.
While Li~\etal~\cite{Li2020} demonstrated the effectiveness of repairing conversational break downs with visual output and touch input, blending in additional modalities complicates the design problems, as designers need to evaluate the integration of a wide variety of visual elements for information input and output.
For example, our early work demonstrates the implementation of Ecological Momentary Assessment (EMA) questionnaires for older adults on a variety of IVA-powered user-attached and user-detached devices that could result in different experiences~\cite{Chen2021}.

\presub
\subsection{Design for Health: Addressing Lack of Support to Access and Manage Health Data}~\label{sec::discussion::features}\postsub
Unlocking the ability for older adults to interact with health data through IVAs could address the frustrations introduced by medication management~(\textbf{B-A1},  \textbf{B-A2}) and health data reporting (\textbf{B-C1}).
Participants reported inefficient and complicated interaction while using today's GUI-based PPs.
This led older adults to stick to more traditional approaches like using pen and paper to keep the medication routines, call providers to report the data, \etc

While integrating voice-first interfaces with existing EHR infrastructure is promising, this does not mean that all features offered by PPs should be addressed through such integration.
Participants (both clinicians and patients) have brought up explicitly barriers under the Daily Life and Routines category (specifically \textbf{B-B2} and \textbf{B-B3} in Table~\ref{tab::findings}), and two major applications: the need of medication reminders, and more support for health data reporting. 
Routinely check-ins, \aka~Ecological Momentary Assessment (EMA), have also been identified as an important need from the providers' perspective.

At a more general level, our findings show the importance of integrating voice into \textbf{(1)} the kind of tasks that happens periodically, and \textbf{(2)} to do that in a way that does not require users to spend significant time and effort.
This echoes Intille~\etal's work~\cite{Intille2016} that uses the concept of microinteractions~\cite{Ashbrook2010} to design an EMA application on smartwatches to increase compliance and completion rate, and reduce the perceived burden on the patients' side.
We believe that the nature of the voice modality, allows for a better design of such micro-interactive task and will help in particular by reducing the devices' access time~\cite{Ashbrook2010}.

Our study also outlined the importance of personalization, when we attempt to design for increased access to health data for older adults.
As noted by C2 and C3, providers stressed the importance of considering the difference between those in their seventies who do not have significant medical issues, and those in their eighties and beyond who start having problems with ability and memory.
Thus the design of IVA systems cannot be one-size-fits-all.
Putting unnecessary features into IVAs would make the system more cumbersome, which would lead to similar frustrations and inefficiencies as the ones caused by PPs (\textbf{B-C3}), likely preventing the successful adoption of IVAs in older adults populations.

\begin{table*}[t]
    \centering
    \includegraphics[width=\textwidth]{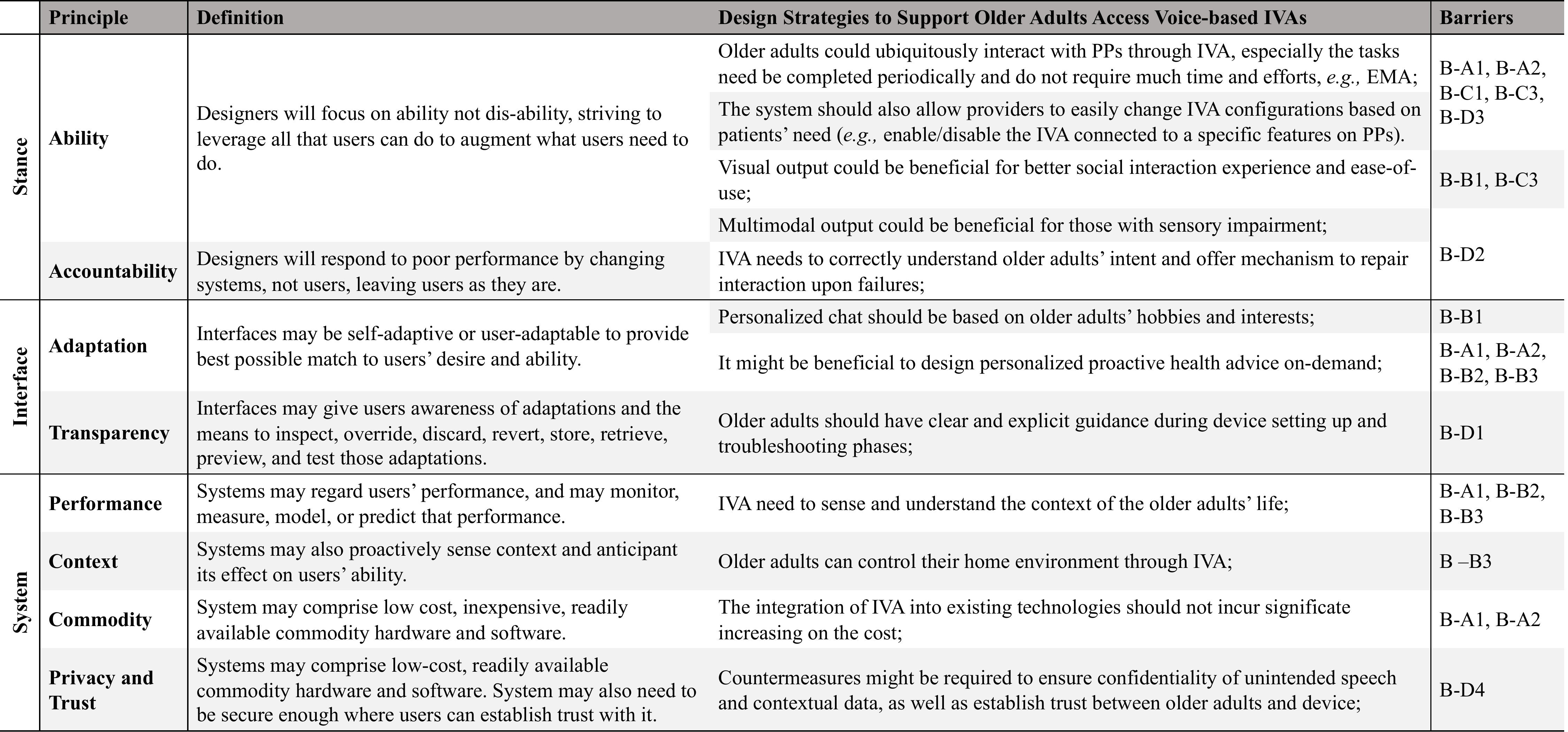}
    \Description{The seven design principles from Ability-based design~\cite{Wobbrock2011, Wobbrock2018} (and additional Privacy and Trust considerations) and how they have been mapped to older adults' needs from our analysis.}
    \captionof{table}{The seven design principles from Ability-based design~\cite{Wobbrock2011, Wobbrock2018} (and additional Privacy and Trust considerations) and how they have been mapped to older adults' needs from our analysis.}
    \label{tab::needs}
    \vspace{-1.0cm}
\end{table*}

\presub
\subsection{Design for Environments: Supporting Ubiquitous Connections to the Environment through Voice}~\label{sec::discussion::env}\postsub
Both providers and patient participants desired being able to use technology to better connect to their environment to perform the right action (\ie~contextual awareness). 
Older adults also mentioned the increased capability to control their surroundings through their voice (\ie~actuation through voice, especially for those with mild mobility impairment, as this feature help them avoid the limb movements to access the control panels, as noted by P9). 
Both providers and patients agreed on striving to achieve consistent monitoring contexts that might cause life-threatening dangers (\eg~detections of falls and giving instructions during extreme medical emergencies) (\textbf{B-B3}). Providers also expect to be able to monitor unhealthy lifestyles (\eg~reminders of regular exercise and warnings to prevent excessive alcohol consumption) (\textbf{B-B2}).
To achieve these goals, it might be necessary to integrate additional sensors and actuators, which could however impede ease-of-use, especially during the stage of device set-up and troubleshooting, one of the important problems reported by our participants (\textbf{B-D1}). It is also important to understand what (and how to) convey the additional information required to operate these sensors and actuators to older adults with voice (or voice-first) modality, especially during the device set-up and error troubleshooting phases.

We believe that user-attached devices with voice-first interface can mitigate this problem by exploiting the sensors components that are already integrated in the devices, and by using additional visual output and touch input to simplify the device set-up phase.
As noted by C2, some providers are trying to use human-aware wearables (\eg~Fitbit), to measure vital signs, notify patients, and remind them of daily routines (\eg~{\it ``time to eat''}\enspace if they have diabetes and want to avoid running low on blood sugar).
As for user-detached (i.e., standalone) IVAs, it is harder to achieve the same effects, especially when using devices with voice-only user interface, where setting up and troubleshooting would lead to the same problems (\textbf{B-D1}).

One promising solution is to realize the concept of ``general purpose sensing,'' which advocates for the use of single sensors to detect multiple and multi-order events~\cite{Gierad2017, Yuvraj2019}. This approach shows potential to simplify setups and operating difficulties and promises to be easily integrated in older adults' everyday life.
Although additional sensors and actuator could help mitigating some older adults’ barriers, they introduce potential issues related to security and sharing of private data, especially when it comes to the proactive conversation triggered by a particular context, a concern that has also been voiced by our participants (\textbf{B-D4}). Privacy concerns
are compounded by additional requirements that would be needed to realize contextually-aware IVAs in the setting of proactive conversation triggered by a specific contextual event.
Today's voice based IVA only supports reactive conversation where a wake words (\eg~``Alexa!'') is required to be announced, to start the voice app.
Instead of consistently streaming user's voice to the cloud, the recognition of the wake word occurs locally on an AI-chip, aiming to prevent unintended speech being shared to the cloud~\cite{AlexaPrivacy}.
However, to create novel solutions that can preemptively alert older adults of any upcoming risk, the system would need to continue streaming contextual data to a remote cloud for understanding users' behavior or speech patterns.

One possible way to resolve this problem is to add additional security countermeasures to identify private speech locally and prevent it to ever reach the microphones.
For example, a recent design of MicShield~\cite{Ke2020} provides a viable solution using inaudible ultrasound to obfuscate and prevent unintended private speech reaching the always-on-and-always-listening microphones.

\presub
\subsection{Design for Abilities: Reframing the Design of Conversational Voice Based IVAs for Older Adults}~\label{sec::discuss:ability}\postsub
We want to close with an additional lens that we would like the research community to use to understand needs and inform the design of IVAs for older adults. We recommend using \textit{ability-based design}~\cite{Wobbrock2011, Wobbrock2018} as a framework to focus on older adults' \textit{ability}, instead of \textit{dis-ability}, when designing interactive systems (\eg~transfer the existing technical skills to IVA use). 
In order to better understand how our work can be situated within the ability-based design framework, we show in Table~\ref{tab::needs} an example of our design strategies to support older adults' access to voice-based IVAs, along with the barriers that emerged from our analysis, and contextualize them under the seven principles of Ability-based design~\cite{Wobbrock2011, Wobbrock2018}. Due to the invasive nature of audio data, and the privacy concerns related to current IVAs, we added \textit{Privacy and Trust} as an additional category.

\presec\section{Limitations and Future Work}~\label{sec::limit}\postsec
\noindent{\bf Participant Recruitment.}
Older adults have a wide range of abilities and experiences, which might not be fully captured by our study.
Therefore, one thread of future work is to focus on diversifying recruited participants.
\First,~the majority of older adult participants either had some fundamental knowledge of general computing devices, or family members who were tech-savvy, which was probably due to the relatively high socioeconomic status before retirement in this particular sample~\cite{SDGDP}.
Older adults living in different situations, without past experience using IVAs, and no access to skilled family members, might have different needs that did not emerge from our study.
Therefore, future studies need to include participants whose inner circle has had less exposure to technology and computing devices.
\Second,~our recruited older adults patients only included those living independently.
Our current findings are, therefore, biased by potential barriers, needs, and IVA use while being supported by caregivers, family, or friends. Including dependent living older adults will open up new barriers, and should be part of future work.
\Finally, we only consider the individual use of IVA on the patients' side. 
Multiple use scenarios, \eg~co-use of user detached IVA and providers' involvements would be considered as future efforts.

\vspace{.5em}
\noindent{\bf Method.} 
In our study, all interviews were conducted individually. 
We expected to hold multiple co-design workshops with patients, care providers, and engage all participants in a real design thinking study.
Typically, user-centered design calls for co-design workshops that are used as the space for ``creative collaboration'', encouraging participants to share stories and coming up with interesting needs~\cite{Sanders2008}.
Due to COVID-19, we had to address challenges of virtual engagement and remote co-design including frequent distractions, unnatural conversation forms, and limited interactions~\cite{Jing2020, Yarmand2021}. 
We believe that in-person co-design workshops might elicit more insights on supporting older adults' healthcare needs through IVAs.

\presec\section{Conclusion}~\label{sec::conclusion}\postsec
This paper presents a need-finding analysis focused on voice based conversational IVAs for older adults aimed at improving QOL and healthcare management.
We report findings using a user-centered design approach, informed by remote and in-person semi--structured interviews with $16$ older adult patients and $5$ healthcare providers respectively, from UC San Diego Health.
By identifying \barriernum~barriers, we unveiled challenges and opportunities for designing effective IVAs for older adults to better manage their health and daily life through the state-of-art information technologies.
Our work will benefit future researchers aiming to create full-fledged IVA software-hardware applications, and to integrate them into existing healthcare systems.

\presec
\begin{acks}\postsec
This work is part of project VOLI\footnote{Project VOLI: \url{http://voli.ucsd.edu}}~\cite{Mrini2021, Chen2021, Johnson2020VoiceBasedCA, Charles2021} and was supported by NIH/NIA under grant R56AG067393.
Co-author Michael Hogarth has an equity interest in LifeLink Inc. and also serves on the company’s Scientific Advisory Board. The terms of this arrangement have been reviewed and approved by the UC San Diego in accordance with its conflict of interest policies.
We appreciate insightful feedback from the anonymous reviewers, and Danilo Gasques, Ru Wang, Khalil Mrini, Matin Yarmand, Thomas Sharkey, and fellow colleagues from the Design Lab and Computer Science and Engineering at UC San Diego.
\end{acks}

\balance
\bibliographystyle{ACM-Reference-Format}
\bibliography{references}


\begin{thebibliography}{92}


\ifx \showCODEN    \undefined \def \showCODEN     #1{\unskip}     \fi
\ifx \showDOI      \undefined \def \showDOI       #1{#1}\fi
\ifx \showISBNx    \undefined \def \showISBNx     #1{\unskip}     \fi
\ifx \showISBNxiii \undefined \def \showISBNxiii  #1{\unskip}     \fi
\ifx \showISSN     \undefined \def \showISSN      #1{\unskip}     \fi
\ifx \showLCCN     \undefined \def \showLCCN      #1{\unskip}     \fi
\ifx \shownote     \undefined \def \shownote      #1{#1}          \fi
\ifx \showarticletitle \undefined \def \showarticletitle #1{#1}   \fi
\ifx \showURL      \undefined \def \showURL       {\relax}        \fi
\providecommand\bibfield[2]{#2}
\providecommand\bibinfo[2]{#2}
\providecommand\natexlab[1]{#1}
\providecommand\showeprint[2][]{arXiv:#2}

\bibitem[\protect\citeauthoryear{??}{Mil}{2016}]%
        {Miller2016}
 \bibinfo{year}{2016}\natexlab{}.
\newblock \showarticletitle{Primary Care Providers’ Views of Patient Portals:
  Interview Study of Perceived Benefits and Consequences}.
\newblock \bibinfo{journal}{\emph{Journal of Medical Internet Research}}
  (\bibinfo{year}{2016}).
\newblock
\urldef\tempurl%
\url{https://doi.org/10.2196/jmir.4953}
\showDOI{\tempurl}


\bibitem[\protect\citeauthoryear{??}{ale}{2019}]%
        {alexa_marketshare}
 \bibinfo{year}{2019}\natexlab{}.
\newblock \bibinfo{booktitle}{\emph{Alexa Devices Maintain 70\% Market Share in
  U.S. according to survey}}.
\newblock
\urldef\tempurl%
\url{https://marketingland.com/alexa-devices-maintain-70-market-share-in-u-s-according-to-survey-265180}
\showURL{%
\tempurl}


\bibitem[\protect\citeauthoryear{??}{blo}{2020}]%
        {blood_pressure_logger}
 \bibinfo{year}{2020}\natexlab{}.
\newblock \bibinfo{title}{Blood Pressure Logger, Amazon Alexa App}.
\newblock
\newblock
\urldef\tempurl%
\url{https://www.amazon.com/HealthyMe-Complete-Health-Wellness-Tracker/dp/B07H2T2R19}
\showURL{%
\tempurl}


\bibitem[\protect\citeauthoryear{??}{pil}{2020}]%
        {pill_me}
 \bibinfo{year}{2020}\natexlab{}.
\newblock \bibinfo{title}{Pill Me! Amazon Alexa App}.
\newblock
\newblock
\urldef\tempurl%
\url{https://www.amazon.com/JLinn-LLC-Pill-Me/dp/B073SPMS53}
\showURL{%
\tempurl}


\bibitem[\protect\citeauthoryear{??}{Ale}{2021}]%
        {AlexaPrivacy}
 \bibinfo{year}{2021}\natexlab{}.
\newblock \bibinfo{title}{{A}lexa and {E}cho {D}evices are {D}esigned to
  {P}rotect {Y}our {P}rivacy}.
\newblock
\newblock
\urldef\tempurl%
\url{https://www.amazon.com/b/?node=19149155011}
\showURL{%
\tempurl}


\bibitem[\protect\citeauthoryear{Agarwal, Harrison, Laput, Boovaraghavan, Chen,
  Hota, Xiao, and Zhang}{Agarwal et~al\mbox{.}}{2019}]%
        {Yuvraj2019}
\bibfield{author}{\bibinfo{person}{Yuvraj Agarwal},
  \bibinfo{person}{Christopher Harrison}, \bibinfo{person}{Gierad Laput},
  \bibinfo{person}{Sudershan Boovaraghavan}, \bibinfo{person}{Chen Chen},
  \bibinfo{person}{Abhijit Hota}, \bibinfo{person}{Bo~Robert Xiao}, {and}
  \bibinfo{person}{Yang Zhang}.} \bibinfo{year}{U.S. Patent 10436615, Oct.
  2019}\natexlab{}.
\newblock \bibinfo{title}{Virtual Sensor System}.
\newblock
\newblock


\bibitem[\protect\citeauthoryear{AgingCare}{AgingCare}{2020}]%
        {agingCare2020}
\bibfield{author}{\bibinfo{person}{AgingCare}.}
  \bibinfo{year}{2020}\natexlab{}.
\newblock \bibinfo{booktitle}{\emph{Dietary Supplements for the Elderly: More
  is Not Always Better}}.
\newblock
\urldef\tempurl%
\url{https://www.agingcare.com/articles/dietary-supplements-for-seniors-more-is-not-always-better-133854.html}
\showURL{%
\tempurl}


\bibitem[\protect\citeauthoryear{Arning and Ziefle}{Arning and Ziefle}{2007}]%
        {Arning2007}
\bibfield{author}{\bibinfo{person}{Katrin Arning} {and}
  \bibinfo{person}{Martina Ziefle}.} \bibinfo{year}{2007}\natexlab{}.
\newblock \showarticletitle{{U}nderstanding {A}ge {D}ifferences in PDA
  {A}cceptance and {P}erformance}.
\newblock \bibinfo{journal}{\emph{Computers in Human Behavior}}
  \bibinfo{volume}{23}, \bibinfo{number}{6} (\bibinfo{year}{2007}),
  \bibinfo{pages}{2904 -- 2927}.
\newblock
\showISSN{0747-5632}
\urldef\tempurl%
\url{https://doi.org/10.1016/j.chb.2006.06.005}
\showDOI{\tempurl}
\newblock
\shownote{Including the Special Issue: Education and Pedagogy with Learning
  Objects and Learning Designs.}


\bibitem[\protect\citeauthoryear{Ashbrook}{Ashbrook}{2010}]%
        {Ashbrook2010}
\bibfield{author}{\bibinfo{person}{Daniel~L. Ashbrook}.}
  \bibinfo{year}{2010}\natexlab{}.
\newblock \emph{\bibinfo{title}{Enabling Mobile Microinteractions}}.
\newblock \bibinfo{thesistype}{Ph.D. Dissertation}. \bibinfo{address}{USA}.
\newblock Advisor(s) Starner, Thad E.
\newblock
\showISBNx{9781124076973}
\newblock
\shownote{AAI3414437.}


\bibitem[\protect\citeauthoryear{Bonilla and Martin-Hammond}{Bonilla and
  Martin-Hammond}{2020}]%
        {Karen2020}
\bibfield{author}{\bibinfo{person}{Karen Bonilla} {and}
  \bibinfo{person}{Aqueasha Martin-Hammond}.} \bibinfo{year}{2020}\natexlab{}.
\newblock \showarticletitle{Older Adults{\textquoteright} Perceptions of
  Intelligent Voice Assistant Privacy, Transparency, and Online Privacy
  Guidelines}. \bibinfo{publisher}{{USENIX} Association}.
\newblock
\urldef\tempurl%
\url{https://www.usenix.org/conference/soups2020/presentation/bonilla}
\showURL{%
\tempurl}


\bibitem[\protect\citeauthoryear{Bosworth, Granger, Mendys, Brindis,
  Burkholder, Czajkowski, Daniel, Ekman, Ho, Johnson, Kimmel, Liu, Musaus,
  Shrank, Buono, Weiss, and Granger}{Bosworth et~al\mbox{.}}{2011}]%
        {Bosworth2011}
\bibfield{author}{\bibinfo{person}{Hayden~B. Bosworth},
  \bibinfo{person}{Bradi~B. Granger}, \bibinfo{person}{Phil Mendys},
  \bibinfo{person}{Ralph Brindis}, \bibinfo{person}{Rebecca Burkholder},
  \bibinfo{person}{Susan~M. Czajkowski}, \bibinfo{person}{Jodi~G. Daniel},
  \bibinfo{person}{Inger Ekman}, \bibinfo{person}{Michael Ho},
  \bibinfo{person}{Mimi Johnson}, \bibinfo{person}{Stephen~E. Kimmel},
  \bibinfo{person}{Larry~Z. Liu}, \bibinfo{person}{John Musaus},
  \bibinfo{person}{William~H. Shrank}, \bibinfo{person}{Elizabeth~Whalley
  Buono}, \bibinfo{person}{Karen Weiss}, {and} \bibinfo{person}{Christopher~B.
  Granger}.} \bibinfo{year}{2011}\natexlab{}.
\newblock \showarticletitle{Medication adherence: A call for action}.
\newblock \bibinfo{journal}{\emph{American Heart Journal}}
  \bibinfo{volume}{162}, \bibinfo{number}{3} (\bibinfo{year}{2011}),
  \bibinfo{pages}{412 -- 424}.
\newblock
\showISSN{0002-8703}
\urldef\tempurl%
\url{https://doi.org/10.1016/j.ahj.2011.06.007}
\showDOI{\tempurl}


\bibitem[\protect\citeauthoryear{Bérubé, Schachner, Keller, Fleisch,
  Wangenheim, Barata, and Kowatsch}{Bérubé et~al\mbox{.}}{2020}]%
        {Burebe2020}
\bibfield{author}{\bibinfo{person}{Caterina Bérubé}, \bibinfo{person}{Theresa
  Schachner}, \bibinfo{person}{Roman Keller}, \bibinfo{person}{Elgar Fleisch},
  \bibinfo{person}{Florian~V. Wangenheim}, \bibinfo{person}{Filipe Barata},
  {and} \bibinfo{person}{Tobias Kowatsch}.} \bibinfo{year}{2020}\natexlab{}.
\newblock \showarticletitle{Voice-based Conversational Agents for the
  Prevention and Management of Chronic and Mental Conditions: A Systematic
  Literature Review}.
\newblock \bibinfo{journal}{\emph{Journal of Medical Internet Research}}
  (\bibinfo{year}{2020}).
\newblock
\urldef\tempurl%
\url{https://www.jmir.org/preprint/25933}
\showURL{%
\tempurl}


\bibitem[\protect\citeauthoryear{Chamberlain, Sharp, and Maiden}{Chamberlain
  et~al\mbox{.}}{2006}]%
        {Stephanie2006}
\bibfield{author}{\bibinfo{person}{Stephanie Chamberlain},
  \bibinfo{person}{Helen Sharp}, {and} \bibinfo{person}{Neil Maiden}.}
  \bibinfo{year}{2006}\natexlab{}.
\newblock \showarticletitle{Towards a Framework for Integrating Agile
  Development and User-Centred Design}. In \bibinfo{booktitle}{\emph{Extreme
  Programming and Agile Processes in Software Engineering}},
  \bibfield{editor}{\bibinfo{person}{Pekka Abrahamsson},
  \bibinfo{person}{Michele Marchesi}, {and} \bibinfo{person}{Giancarlo Succi}}
  (Eds.). \bibinfo{publisher}{Springer Berlin Heidelberg},
  \bibinfo{address}{Berlin, Heidelberg}, \bibinfo{pages}{143 -- 153}.
\newblock
\showISBNx{978-3-540-35095-8}


\bibitem[\protect\citeauthoryear{Chandel, Devanuj, and Doke}{Chandel
  et~al\mbox{.}}{2013}]%
        {Chandel2013}
\bibfield{author}{\bibinfo{person}{Priyanka Chandel},
  \bibinfo{person}{Devanuj}, {and} \bibinfo{person}{Pankaj Doke}.}
  \bibinfo{year}{2013}\natexlab{}.
\newblock \showarticletitle{A Comparative Study of Voice and Graphical User
  Interfaces with Respect to Literacy Levels}. In
  \bibinfo{booktitle}{\emph{Proceedings of the 3rd ACM Symposium on Computing
  for Development}} (Bangalore, India) \emph{(\bibinfo{series}{ACM DEV '13})}.
  \bibinfo{publisher}{Association for Computing Machinery},
  \bibinfo{address}{New York, NY, USA}, Article \bibinfo{articleno}{33},
  \bibinfo{numpages}{2}~pages.
\newblock
\showISBNx{9781450318563}
\urldef\tempurl%
\url{https://doi.org/10.1145/2442882.2442921}
\showDOI{\tempurl}


\bibitem[\protect\citeauthoryear{Charles, Chen, Johnson, Lee, Lifset, Hogarth,
  Weibel, Farcas, and Moore}{Charles et~al\mbox{.}}{2021}]%
        {Charles2021}
\bibfield{author}{\bibinfo{person}{Kemeberley Charles}, \bibinfo{person}{Chen
  Chen}, \bibinfo{person}{Janet~G. Johnson}, \bibinfo{person}{Alice Lee},
  \bibinfo{person}{Ella~T. Lifset}, \bibinfo{person}{Michael Hogarth},
  \bibinfo{person}{Nadir Weibel}, \bibinfo{person}{Emilia Farcas}, {and}
  \bibinfo{person}{Alison~A. Moore}.} \bibinfo{year}{2021}\natexlab{}.
\newblock \showarticletitle{{H}ow {M}ight an {I}ntelligent {V}oice {A}ssistant
  {A}ddress {O}lder {A}dults' {H}ealth-{R}elated {N}eeds?}. In
  \bibinfo{booktitle}{\emph{Journal of the American Geriatrics Society}},
  Vol.~\bibinfo{volume}{69}. Willey 111 River Street, Hoboken 07030-5774, NJ,
  USA, \bibinfo{pages}{S243--S244}.
\newblock


\bibitem[\protect\citeauthoryear{Chen, Mrini, Charles, Lifset, Hogarth, Moore,
  Weibel, and Farcas}{Chen et~al\mbox{.}}{2021}]%
        {Chen2021}
\bibfield{author}{\bibinfo{person}{Chen Chen}, \bibinfo{person}{Khalil Mrini},
  \bibinfo{person}{Kemeberly Charles}, \bibinfo{person}{Ella~T. Lifset},
  \bibinfo{person}{Michael Hogarth}, \bibinfo{person}{Alison~A. Moore},
  \bibinfo{person}{Nadir Weibel}, {and} \bibinfo{person}{Emilia Farcas}.}
  \bibinfo{year}{2021}\natexlab{}.
\newblock \showarticletitle{{T}oward a {U}nified {M}etadata {S}chema for
  {E}cological {M}omentary {A}ssessment with {V}oice -- {F}irst {V}irtual
  {A}ssistants}. In \bibinfo{booktitle}{\emph{Proceedings of the 3rd Conference
  on Conversational User Interfaces}} (Bilbao (online), AA, Spain)
  \emph{(\bibinfo{series}{CUI '21})}. \bibinfo{publisher}{Association for
  Computing Machinery}, \bibinfo{address}{Bilbao (online), AA, Spain}, 6.
\newblock
\showISBNx{9781450389983}
\urldef\tempurl%
\url{https://doi.org/10.1145/3469595.3469626}
\showDOI{\tempurl}


\bibitem[\protect\citeauthoryear{Czaja, Boot, Charness, and Rogers}{Czaja
  et~al\mbox{.}}{2019}]%
        {Czaja2019}
\bibfield{author}{\bibinfo{person}{Sara~J. Czaja}, \bibinfo{person}{Walter~R.
  Boot}, \bibinfo{person}{Neil Charness}, {and} \bibinfo{person}{Wendy~A.
  Rogers}.} \bibinfo{year}{2019}\natexlab{}.
\newblock \bibinfo{booktitle}{\emph{Designing for Older Adults: Principles and
  Creative Human Factors Approaches} (\bibinfo{edition}{3} ed.)}.
\newblock \bibinfo{publisher}{CRC Press}.
\newblock
\showISBNx{9781138053663}


\bibitem[\protect\citeauthoryear{Davis, Bagozzi, and Warshaw}{Davis
  et~al\mbox{.}}{[n.d.]}]%
        {Davis1989}
\bibfield{author}{\bibinfo{person}{Fred~D. Davis}, \bibinfo{person}{Richard~P.
  Bagozzi}, {and} \bibinfo{person}{Paul~R. Warshaw}.}
  \bibinfo{year}{[n.d.]}\natexlab{}.
\newblock \showarticletitle{User Acceptance of Computer Technology: A
  Comparison of Two Theoretical Models}.
\newblock  \bibinfo{volume}{35}, \bibinfo{number}{8}
  (\bibinfo{year}{[n.\,d.]}), \bibinfo{pages}{982--1003}.
\newblock
\showISSN{00251909, 15265501}
\urldef\tempurl%
\url{http://www.jstor.org/stable/2632151}
\showURL{%
\tempurl}
\newblock
\shownote{Publisher: {INFORMS}.}


\bibitem[\protect\citeauthoryear{Demakakos, Nunn, and Nazroo}{Demakakos
  et~al\mbox{.}}{2006}]%
        {Demakakos2006}
\bibfield{author}{\bibinfo{person}{Panayotes Demakakos}, \bibinfo{person}{Susan
  Nunn}, {and} \bibinfo{person}{James Nazroo}.}
  \bibinfo{year}{2006}\natexlab{}.
\newblock \showarticletitle{Loneliness, Relative Deprivation and Life
  Satisfaction}.
\newblock


\bibitem[\protect\citeauthoryear{Doyle, Walsh, Sassu, and McDonagh}{Doyle
  et~al\mbox{.}}{2014}]%
        {Doyle2014}
\bibfield{author}{\bibinfo{person}{Julie Doyle}, \bibinfo{person}{Lorcan
  Walsh}, \bibinfo{person}{Antonella Sassu}, {and} \bibinfo{person}{Teresa
  McDonagh}.} \bibinfo{year}{2014}\natexlab{}.
\newblock \showarticletitle{Designing a Wellness Self-Management Tool for Older
  Adults: Results from a Field Trial of YourWellness}. In
  \bibinfo{booktitle}{\emph{Proceedings of the 8th International Conference on
  Pervasive Computing Technologies for Healthcare}} (Oldenburg, Germany)
  \emph{(\bibinfo{series}{PervasiveHealth '14})}. \bibinfo{publisher}{ICST
  (Institute for Computer Sciences, Social-Informatics and Telecommunications
  Engineering)}, \bibinfo{address}{Brussels, BEL}, \bibinfo{pages}{134–141}.
\newblock
\showISBNx{9781631900112}
\urldef\tempurl%
\url{https://doi.org/10.4108/icst.pervasivehealth.2014.254950}
\showDOI{\tempurl}


\bibitem[\protect\citeauthoryear{Elo and Kyng{\"a}s}{Elo and
  Kyng{\"a}s}{2008}]%
        {Satu2008}
\bibfield{author}{\bibinfo{person}{Satu Elo} {and} \bibinfo{person}{Helvi
  Kyng{\"a}s}.} \bibinfo{year}{2008}\natexlab{}.
\newblock \showarticletitle{The Qualitative Content Analysis Process}.
\newblock \bibinfo{journal}{\emph{Journal of Advanced Nursing}}
  \bibinfo{volume}{62}, \bibinfo{number}{1} (\bibinfo{year}{2008}),
  \bibinfo{pages}{107 -- 115}.
\newblock
\urldef\tempurl%
\url{https://doi.org/10.1111/j.1365-2648.2007.04569.x}
\showDOI{\tempurl}


\bibitem[\protect\citeauthoryear{Federman, Sarzynski, Brach, Francaviglia,
  Jacques, Jandorf, Munoz, Wolf, and Kannry}{Federman et~al\mbox{.}}{2018}]%
        {Federman2018}
\bibfield{author}{\bibinfo{person}{Alex Federman}, \bibinfo{person}{Erin
  Sarzynski}, \bibinfo{person}{Cindy Brach}, \bibinfo{person}{Paul
  Francaviglia}, \bibinfo{person}{Jessica Jacques}, \bibinfo{person}{Lina
  Jandorf}, \bibinfo{person}{Angela~Sanchez Munoz}, \bibinfo{person}{Michael
  Wolf}, {and} \bibinfo{person}{Joseph Kannry}.}
  \bibinfo{year}{2018}\natexlab{}.
\newblock \showarticletitle{{C}hallenges {O}ptimizing the {A}fter {V}isit
  {S}ummary}.
\newblock \bibinfo{journal}{\emph{Journal of Medical Internet Research}}
  (\bibinfo{date}{15 Sep} \bibinfo{year}{2018}).
\newblock
\urldef\tempurl%
\url{https://doi.org/10.1016/j.ijmedinf.2018.09.009}
\showDOI{\tempurl}


\bibitem[\protect\citeauthoryear{Ferreira, Almeida, Rosa, Oliveira, Teixeira,
  and Pereira}{Ferreira et~al\mbox{.}}{2013}]%
        {Ferreira2013}
\bibfield{author}{\bibinfo{person}{Flávio Ferreira}, \bibinfo{person}{Nuno
  Almeida}, \bibinfo{person}{Ana~Filipa Rosa}, \bibinfo{person}{André
  Oliveira}, \bibinfo{person}{António Teixeira}, {and}
  \bibinfo{person}{José~Casimiro Pereira}.} \bibinfo{year}{2013}\natexlab{}.
\newblock \showarticletitle{Multimodal and {A}daptable {M}edication {A}ssistant
  for the {E}lderly: {A} {P}rototype for {I}nteraction and {U}sability in
  {S}martphones}. In \bibinfo{booktitle}{\emph{2013 8th Iberian Conference on
  Information Systems and Technologies (CISTI)}}. \bibinfo{pages}{1--6}.
\newblock


\bibitem[\protect\citeauthoryear{for Diseases~Control and (CDC)}{for
  Diseases~Control and (CDC)}{2020}]%
        {retirement_community_covid19}
\bibfield{author}{\bibinfo{person}{Centers for Diseases~Control} {and}
  \bibinfo{person}{Prevention (CDC)}.} \bibinfo{year}{2020}\natexlab{}.
\newblock \bibinfo{booktitle}{\emph{Considerations for Retirement Communities
  and Independent Living Facilities}}.
\newblock
\urldef\tempurl%
\url{https://www.cdc.gov/coronavirus/2019-ncov/community/retirement/considerations.html}
\showURL{%
\tempurl}


\bibitem[\protect\citeauthoryear{for Quality Assurance~(NCQA)}{for Quality
  Assurance~(NCQA)}{2020}]%
        {ncqa_health_quality_guide}
\bibfield{author}{\bibinfo{person}{National~Committee for Quality
  Assurance~(NCQA)}.} \bibinfo{year}{2020}\natexlab{}.
\newblock \showarticletitle{The Essential Guide to Health Care Quality}.
\newblock  (\bibinfo{year}{2020}).
\newblock
\urldef\tempurl%
\url{https://www.kdheks.gov/hcf/news/download/04172007_ncqa_health_quality.pdf}
\showURL{%
\tempurl}


\bibitem[\protect\citeauthoryear{Garrett, Heller, Fowler, Alberto, Fredrick,
  and O{\textquoteright}Rourke}{Garrett et~al\mbox{.}}{2011}]%
        {Jennifer2011}
\bibfield{author}{\bibinfo{person}{Jennifer~Tumlin Garrett},
  \bibinfo{person}{Kathryn~Wolff Heller}, \bibinfo{person}{Linda~P. Fowler},
  \bibinfo{person}{Paul~A. Alberto}, \bibinfo{person}{Laura~D. Fredrick}, {and}
  \bibinfo{person}{Colleen~M. O{\textquoteright}Rourke}.}
  \bibinfo{year}{2011}\natexlab{}.
\newblock \showarticletitle{Using Speech Recognition Software to Increase
  Writing Fluency for Individuals with Physical Disabilities}.
\newblock \bibinfo{journal}{\emph{Journal of Special Education Technology}}
  \bibinfo{volume}{26}, \bibinfo{number}{1} (\bibinfo{year}{2011}),
  \bibinfo{pages}{25--41}.
\newblock
\urldef\tempurl%
\url{https://doi.org/10.1177/016264341102600104}
\showDOI{\tempurl}


\bibitem[\protect\citeauthoryear{Goldzweig, Orshansky, Paige, Towfigh,
  Haggstrom, Miake-Lye, Beroes, and Shekelle}{Goldzweig et~al\mbox{.}}{2013}]%
        {Goldzweig2013}
\bibfield{author}{\bibinfo{person}{Caroline~Lubick Goldzweig},
  \bibinfo{person}{Greg Orshansky}, \bibinfo{person}{Neil~M. Paige},
  \bibinfo{person}{Ali~Alexander Towfigh}, \bibinfo{person}{David~A.
  Haggstrom}, \bibinfo{person}{Isomi Miake-Lye}, \bibinfo{person}{Jessica~M.
  Beroes}, {and} \bibinfo{person}{Paul~G. Shekelle}.}
  \bibinfo{year}{2013}\natexlab{}.
\newblock \showarticletitle{{E}lectronic {P}atient {P}ortals: {E}vidence on
  {H}ealth {O}utcomes, {S}atisfaction, {E}fficiency, and {A}ttitudes: a
  {S}ystematic {R}eview}.
\newblock \bibinfo{journal}{\emph{Annals of Internal Medicine}}
  \bibinfo{volume}{159}, \bibinfo{number}{10} (\bibinfo{date}{Nov}
  \bibinfo{year}{2013}), \bibinfo{pages}{677--687}.
\newblock
\urldef\tempurl%
\url{https://doi.org/10.7326/0003-4819-159-10-201311190-00006}
\showDOI{\tempurl}


\bibitem[\protect\citeauthoryear{He}{He}{2020}]%
        {Jing2020}
\bibfield{author}{\bibinfo{person}{Jing He}.} \bibinfo{year}{2020}\natexlab{}.
\newblock \emph{\bibinfo{title}{Senior Related IxD Research and Remote
  Co-Design with Seniors - A Literature Review and Exploratory Study}}.
\newblock \bibinfo{thesistype}{Master's\ thesis}. \bibinfo{school}{Department
  of Computer Science, Aalborg University}, \bibinfo{address}{Fredrik Bajers
  Vej 7K, 9220 Aalborg East, Denmark}.
\newblock


\bibitem[\protect\citeauthoryear{Heinz, Martin, Margrett, Yearns, Franke, Yang,
  Wong, and Chang}{Heinz et~al\mbox{.}}{2013}]%
        {Heinz2013}
\bibfield{author}{\bibinfo{person}{Melinda Heinz}, \bibinfo{person}{Peter
  Martin}, \bibinfo{person}{Jennifer~A. Margrett}, \bibinfo{person}{Mary
  Yearns}, \bibinfo{person}{Warren Franke}, \bibinfo{person}{Hen-I Yang},
  \bibinfo{person}{Johnny Wong}, {and} \bibinfo{person}{Carl~K. Chang}.}
  \bibinfo{year}{2013}\natexlab{}.
\newblock \showarticletitle{{P}erceptions of {T}echnology among {O}lder
  {A}dults}.
\newblock \bibinfo{journal}{\emph{J Gerontol Nurs}} \bibinfo{volume}{39},
  \bibinfo{number}{1} (\bibinfo{date}{Jan} \bibinfo{year}{2013}),
  \bibinfo{pages}{42--51}.
\newblock
\urldef\tempurl%
\url{https://doi.org/10.3928/00989134-20121204-04}
\showDOI{\tempurl}


\bibitem[\protect\citeauthoryear{House}{House}{2020}]%
        {covid19}
\bibfield{author}{\bibinfo{person}{The~White House}.}
  \bibinfo{year}{2020}\natexlab{}.
\newblock \bibinfo{booktitle}{\emph{Proclamation on Declaring a National
  Emergency Concerning the Novel Coronavirus Disease ({COVID}-19) Outbreak}}.
\newblock
\urldef\tempurl%
\url{https://www.whitehouse.gov/presidential-actions/proclamation-declaring-national-emergency-concerning-novel-coronavirus-disease-covid-19-outbreak/}
\showURL{%
\tempurl}


\bibitem[\protect\citeauthoryear{Hoy}{Hoy}{2018}]%
        {Hoy2018}
\bibfield{author}{\bibinfo{person}{Matthew Hoy}.}
  \bibinfo{year}{2018}\natexlab{}.
\newblock \showarticletitle{Alexa, Siri, Cortana, and More: An Introduction to
  Voice Assistants}.
\newblock \bibinfo{journal}{\emph{Medical Reference Services Quarterly}}
  \bibinfo{volume}{37} (\bibinfo{date}{01} \bibinfo{year}{2018}),
  \bibinfo{pages}{81--88}.
\newblock
\urldef\tempurl%
\url{https://doi.org/10.1080/02763869.2018.1404391}
\showDOI{\tempurl}


\bibitem[\protect\citeauthoryear{Intille, Haynes, Maniar, Ponnada, and
  Manjourides}{Intille et~al\mbox{.}}{2016}]%
        {Intille2016}
\bibfield{author}{\bibinfo{person}{Stephen Intille}, \bibinfo{person}{Caitlin
  Haynes}, \bibinfo{person}{Dharam Maniar}, \bibinfo{person}{Aditya Ponnada},
  {and} \bibinfo{person}{Justin Manjourides}.} \bibinfo{year}{2016}\natexlab{}.
\newblock \showarticletitle{$\mu$EMA: Microinteraction-Based Ecological
  Momentary Assessment (EMA) Using a Smartwatch}. In
  \bibinfo{booktitle}{\emph{Proceedings of the 2016 ACM International Joint
  Conference on Pervasive and Ubiquitous Computing}} (Heidelberg, Germany)
  \emph{(\bibinfo{series}{UbiComp '16})}. \bibinfo{publisher}{Association for
  Computing Machinery}, \bibinfo{address}{New York, NY, USA},
  \bibinfo{pages}{1124–1128}.
\newblock
\showISBNx{9781450344616}
\urldef\tempurl%
\url{https://doi.org/10.1145/2971648.2971717}
\showDOI{\tempurl}


\bibitem[\protect\citeauthoryear{Irizarry, Dabbs, and Curran}{Irizarry
  et~al\mbox{.}}{2015}]%
        {Taya2015}
\bibfield{author}{\bibinfo{person}{Taya Irizarry},
  \bibinfo{person}{Annette~DeVito Dabbs}, {and} \bibinfo{person}{Christine~R
  Curran}.} \bibinfo{year}{2015}\natexlab{}.
\newblock \showarticletitle{Patient portals and patient engagement: a state of
  the science review}.
\newblock \bibinfo{journal}{\emph{Journal of Medical Internet Research}}
  \bibinfo{volume}{17}, \bibinfo{number}{6} (\bibinfo{year}{2015}),
  \bibinfo{pages}{e148}.
\newblock
\urldef\tempurl%
\url{https://doi.org/10.2196/jmir.4255}
\showDOI{\tempurl}


\bibitem[\protect\citeauthoryear{Johnson, Mrini, Hogarth, Moore, Nakashole,
  Weibel, and Farcas}{Johnson et~al\mbox{.}}{2020}]%
        {Johnson2020VoiceBasedCA}
\bibfield{author}{\bibinfo{person}{Janet~G. Johnson}, \bibinfo{person}{Khalil
  Mrini}, \bibinfo{person}{Michael Hogarth}, \bibinfo{person}{Alison~A. Moore},
  \bibinfo{person}{Ndapa Nakashole}, \bibinfo{person}{Nadir Weibel}, {and}
  \bibinfo{person}{Emilia Farcas}.} \bibinfo{year}{2020}\natexlab{}.
\newblock \showarticletitle{{V}oice--{B}ased {C}onversational {A}gents for
  {O}lder {A}dults}. In \bibinfo{booktitle}{\emph{Proceedings of the 2020 CHI
  Conference on Human Factors in Computing Systems Workshop on Conversational
  Agents for Health and Wellbeing}}.
\newblock


\bibitem[\protect\citeauthoryear{Kim}{Kim}{2021}]%
        {Kim2021}
\bibfield{author}{\bibinfo{person}{Sunyoung Kim}.}
  \bibinfo{year}{2021}\natexlab{}.
\newblock \showarticletitle{{E}xploring {H}ow {O}lder {A}dults {U}se a {S}mart
  {S}peaker -- {B}ased {V}oice {A}ssistant in {T}heir {F}irst {I}nteractions:
  {Q}ualitative {S}tudy}.
\newblock \bibinfo{journal}{\emph{JMIR Mhealth Uhealth}} \bibinfo{volume}{9},
  \bibinfo{number}{1} (\bibinfo{date}{13 Jan} \bibinfo{year}{2021}),
  \bibinfo{pages}{e20427}.
\newblock
\showISSN{2291-5222}
\urldef\tempurl%
\url{https://doi.org/10.2196/20427}
\showDOI{\tempurl}


\bibitem[\protect\citeauthoryear{Kim and Fadem}{Kim and Fadem}{2018}]%
        {Kim2018}
\bibfield{author}{\bibinfo{person}{Sunyoung Kim} {and} \bibinfo{person}{Sarah
  Fadem}.} \bibinfo{year}{2018}\natexlab{}.
\newblock \showarticletitle{{C}ommunication {M}atters: {E}xploring {O}lder
  {A}dults' {C}urrent {U}se of {P}atient {P}ortals}.
\newblock \bibinfo{journal}{\emph{International Journal of Medical
  Informatics}}  \bibinfo{volume}{120} (\bibinfo{date}{12}
  \bibinfo{year}{2018}), \bibinfo{pages}{126--136}.
\newblock
\urldef\tempurl%
\url{https://doi.org/10.1016/j.ijmedinf.2018.10.004}
\showDOI{\tempurl}


\bibitem[\protect\citeauthoryear{Kim, Gajos, Muller, and Grosz}{Kim
  et~al\mbox{.}}{2016}]%
        {Kim2016}
\bibfield{author}{\bibinfo{person}{Sunyoung Kim}, \bibinfo{person}{Krzysztof~Z.
  Gajos}, \bibinfo{person}{Michael Muller}, {and} \bibinfo{person}{Barbara~J.
  Grosz}.} \bibinfo{year}{2016}\natexlab{}.
\newblock \showarticletitle{Acceptance of Mobile Technology by Older Adults: A
  Preliminary Study}. In \bibinfo{booktitle}{\emph{Proceedings of the 18th
  International Conference on Human-Computer Interaction with Mobile Devices
  and Services}} (Florence, Italy) \emph{(\bibinfo{series}{MobileHCI '16})}.
  \bibinfo{publisher}{Association for Computing Machinery},
  \bibinfo{address}{New York, NY, USA}, \bibinfo{pages}{147–157}.
\newblock
\showISBNx{9781450344081}
\urldef\tempurl%
\url{https://doi.org/10.1145/2935334.2935380}
\showDOI{\tempurl}


\bibitem[\protect\citeauthoryear{Kinsella}{Kinsella}{2019}]%
        {va_population}
\bibfield{author}{\bibinfo{person}{Bert Kinsella}.}
  \bibinfo{year}{2019}\natexlab{}.
\newblock \bibinfo{booktitle}{\emph{Voice Assistant Demographic Data – Young
  Consumers More Likely to Own Smart Speakers While Over 60 Bias Toward Alexa
  and Siri}}.
\newblock
\urldef\tempurl%
\url{https://voicebot.ai/2019/06/21/voice-assistant-demographic-data-young-consumers-more-likely-to-own-smart-speakers-while-over-60-bias-toward-alexa-and-siri/}
\showURL{%
\tempurl}


\bibitem[\protect\citeauthoryear{Kobayashi, Kosugi, Takagi, Nemoto, Nemoto,
  Arai, and Yamada}{Kobayashi et~al\mbox{.}}{2019}]%
        {Kobayashi2019}
\bibfield{author}{\bibinfo{person}{Masatomo Kobayashi},
  \bibinfo{person}{Akihiro Kosugi}, \bibinfo{person}{Hironobu Takagi},
  \bibinfo{person}{Miyuki Nemoto}, \bibinfo{person}{Kiyotaka Nemoto},
  \bibinfo{person}{Tetsuaki Arai}, {and} \bibinfo{person}{Yasunori Yamada}.}
  \bibinfo{year}{2019}\natexlab{}.
\newblock \showarticletitle{Effects of Age-Related Cognitive Decline on Elderly
  User Interactions with Voice-Based Dialogue Systems}. In
  \bibinfo{booktitle}{\emph{Human-Computer Interaction -- INTERACT 2019}},
  \bibfield{editor}{\bibinfo{person}{David Lamas}, \bibinfo{person}{Fernando
  Loizides}, \bibinfo{person}{Lennart Nacke}, \bibinfo{person}{Helen Petrie},
  \bibinfo{person}{Marco Winckler}, {and} \bibinfo{person}{Panayiotis
  Zaphiris}} (Eds.). \bibinfo{publisher}{Springer International Publishing},
  \bibinfo{address}{Cham}, \bibinfo{pages}{53--74}.
\newblock
\showISBNx{978-3-030-29390-1}


\bibitem[\protect\citeauthoryear{Kononova, Li, Kamp, Bowen, Rikard, Cotten, and
  Peng}{Kononova et~al\mbox{.}}{2019}]%
        {Kononova2019}
\bibfield{author}{\bibinfo{person}{Anastasia Kononova}, \bibinfo{person}{Lin
  Li}, \bibinfo{person}{Kendra Kamp}, \bibinfo{person}{Marie Bowen},
  \bibinfo{person}{R~V Rikard}, \bibinfo{person}{Shelia Cotten}, {and}
  \bibinfo{person}{Wei Peng}.} \bibinfo{year}{2019}\natexlab{}.
\newblock \showarticletitle{The Use of Wearable Activity Trackers Among Older
  Adults: Focus Group Study of Tracker Perceptions, Motivations, and Barriers
  in the Maintenance Stage of Behavior Change}.
\newblock \bibinfo{journal}{\emph{Journal of Medical Internet Research Mhealth
  Uhealth}} \bibinfo{volume}{7}, \bibinfo{number}{4} (\bibinfo{date}{5 4}
  \bibinfo{year}{2019}), \bibinfo{pages}{e9832}.
\newblock
\urldef\tempurl%
\url{https://doi.org/10.2196/mhealth.9832}
\showDOI{\tempurl}


\bibitem[\protect\citeauthoryear{Kowalski, Jaskulska, Skorupska, Abramczuk,
  Biele, Kope\'{c}, and Marasek}{Kowalski et~al\mbox{.}}{2019}]%
        {Jaroslaw2019}
\bibfield{author}{\bibinfo{person}{Jaros\l{}aw Kowalski}, \bibinfo{person}{Anna
  Jaskulska}, \bibinfo{person}{Kinga Skorupska}, \bibinfo{person}{Katarzyna
  Abramczuk}, \bibinfo{person}{Cezary Biele}, \bibinfo{person}{Wies\l{}aw
  Kope\'{c}}, {and} \bibinfo{person}{Krzysztof Marasek}.}
  \bibinfo{year}{2019}\natexlab{}.
\newblock \showarticletitle{Older Adults and Voice Interaction: A Pilot Study
  with Google Home}. In \bibinfo{booktitle}{\emph{Extended Abstracts of the
  2019 CHI Conference on Human Factors in Computing Systems}} (Glasgow,
  Scotland, UK) \emph{(\bibinfo{series}{CHI EA '19})}.
  \bibinfo{publisher}{Association for Computing Machinery},
  \bibinfo{address}{New York, NY, USA}, \bibinfo{pages}{1–6}.
\newblock
\showISBNx{9781450359719}
\urldef\tempurl%
\url{https://doi.org/10.1145/3290607.3312973}
\showDOI{\tempurl}


\bibitem[\protect\citeauthoryear{Kuflinski}{Kuflinski}{2019}]%
        {Yaroslav2019}
\bibfield{author}{\bibinfo{person}{Yaroslav Kuflinski}.}
  \bibinfo{year}{2019}\natexlab{}.
\newblock \bibinfo{booktitle}{\emph{AI-based Smart Speakers in Healthcare: a
  Game Changer or a Fad?}}
\newblock
\urldef\tempurl%
\url{https://www.healtheuropa.eu/ai-based-smart-speakers-in-healthcare/92189/}
\showURL{%
\tempurl}


\bibitem[\protect\citeauthoryear{Kumah-Crystal, Pirtle, Whyte, Goode, Anders,
  and Lehmann}{Kumah-Crystal et~al\mbox{.}}{2018}]%
        {Kumah2018}
\bibfield{author}{\bibinfo{person}{Yaa~A. Kumah-Crystal},
  \bibinfo{person}{Claude~J. Pirtle}, \bibinfo{person}{Harrison~M. Whyte},
  \bibinfo{person}{Edward~S. Goode}, \bibinfo{person}{Shilo~H. Anders}, {and}
  \bibinfo{person}{Christopher~U. Lehmann}.} \bibinfo{year}{2018}\natexlab{}.
\newblock \showarticletitle{{E}lectronic {H}ealth {R}ecord {I}nteractions
  through {V}oice: {A} {R}eview}.
\newblock \bibinfo{journal}{\emph{Appl Clin Inform}} \bibinfo{volume}{9},
  \bibinfo{number}{3} (\bibinfo{date}{Jul} \bibinfo{year}{2018}),
  \bibinfo{pages}{541--552}.
\newblock
\urldef\tempurl%
\url{https://doi.org/10.1055/s-0038-1666844}
\showDOI{\tempurl}


\bibitem[\protect\citeauthoryear{Laput, Zhang, and Harrison}{Laput
  et~al\mbox{.}}{2017}]%
        {Gierad2017}
\bibfield{author}{\bibinfo{person}{Gierad Laput}, \bibinfo{person}{Yang Zhang},
  {and} \bibinfo{person}{Chris Harrison}.} \bibinfo{year}{2017}\natexlab{}.
\newblock \showarticletitle{Synthetic Sensors: Towards General-Purpose
  Sensing}. In \bibinfo{booktitle}{\emph{Proceedings of the 2017 CHI Conference
  on Human Factors in Computing Systems}} (Denver, Colorado, USA)
  \emph{(\bibinfo{series}{CHI '17})}. \bibinfo{publisher}{Association for
  Computing Machinery}, \bibinfo{address}{New York, NY, USA},
  \bibinfo{pages}{3986–3999}.
\newblock
\showISBNx{9781450346559}
\urldef\tempurl%
\url{https://doi.org/10.1145/3025453.3025773}
\showDOI{\tempurl}


\bibitem[\protect\citeauthoryear{Laranjo, Dunn, Tong, Kocaballi, Chen, Bashir,
  Surian, Gallego, Magrabi, Lau, and Coiera}{Laranjo et~al\mbox{.}}{2018}]%
        {Laranjo2018}
\bibfield{author}{\bibinfo{person}{Liliana Laranjo}, \bibinfo{person}{Adam~G
  Dunn}, \bibinfo{person}{Huong~Ly Tong}, \bibinfo{person}{Ahmet~Baki
  Kocaballi}, \bibinfo{person}{Jessica Chen}, \bibinfo{person}{Rabia Bashir},
  \bibinfo{person}{Didi Surian}, \bibinfo{person}{Blanca Gallego},
  \bibinfo{person}{Farah Magrabi}, \bibinfo{person}{Annie Y~S Lau}, {and}
  \bibinfo{person}{Enrico Coiera}.} \bibinfo{year}{2018}\natexlab{}.
\newblock \showarticletitle{Conversational {A}gents in {H}ealthcare: a
  {S}ystematic {R}eview}.
\newblock  \bibinfo{volume}{25}, \bibinfo{number}{9} (\bibinfo{year}{2018}),
  \bibinfo{pages}{1248--1258}.
\newblock
\showISSN{1527-974X}
\urldef\tempurl%
\url{https://doi.org/10.1093/jamia/ocy072}
\showDOI{\tempurl}


\bibitem[\protect\citeauthoryear{Laskowski-Jones}{Laskowski-Jones}{2014}]%
        {noauthorOTC2014}
\bibfield{author}{\bibinfo{person}{Linda Laskowski-Jones}.}
  \bibinfo{year}{2014}\natexlab{}.
\newblock \showarticletitle{{OTC} Supplements: Healthy or Not?}
\newblock \bibinfo{journal}{\emph{Nursing 2014}} \bibinfo{volume}{44},
  \bibinfo{number}{11} (\bibinfo{date}{Nov} \bibinfo{year}{2014}),
  \bibinfo{pages}{6}.
\newblock
\showISSN{0360-4039}
\urldef\tempurl%
\url{https://doi.org/10.1097/01.NURSE.0000454976.04290.fd}
\showDOI{\tempurl}


\bibitem[\protect\citeauthoryear{Latulipe, Quandt, Melius, Bertoni, Miller,
  Smith, and Arcury}{Latulipe et~al\mbox{.}}{2018}]%
        {Latulipe2018}
\bibfield{author}{\bibinfo{person}{Celine Latulipe}, \bibinfo{person}{Sara~A.
  Quandt}, \bibinfo{person}{Kathryn~Altizer Melius}, \bibinfo{person}{Alain
  Bertoni}, \bibinfo{person}{David~P. Miller}, \bibinfo{person}{Douglas Smith},
  {and} \bibinfo{person}{Thomas~A. Arcury}.} \bibinfo{year}{2018}\natexlab{}.
\newblock \showarticletitle{{I}nsights {I}nto {O}lder {A}dult {P}atient
  {C}oncerns {A}round the {C}aregiver {P}roxy {P}ortal {U}se: {Q}ualitative
  {I}nterview {S}tudy}.
\newblock \bibinfo{journal}{\emph{Journal of Medical Internet Research}}
  \bibinfo{volume}{20}, \bibinfo{number}{11} (\bibinfo{date}{Nov}
  \bibinfo{year}{2018}).
\newblock
\urldef\tempurl%
\url{https://doi.org/10.2196/10524}
\showDOI{\tempurl}


\bibitem[\protect\citeauthoryear{Li, Chen, Xia, Mitchell, and Myers}{Li
  et~al\mbox{.}}{2020}]%
        {Li2020}
\bibfield{author}{\bibinfo{person}{Toby Jia-Jun Li}, \bibinfo{person}{Jingya
  Chen}, \bibinfo{person}{Haijun Xia}, \bibinfo{person}{Tom~M. Mitchell}, {and}
  \bibinfo{person}{Brad~A. Myers}.} \bibinfo{year}{2020}\natexlab{}.
\newblock \showarticletitle{Multi-Modal Repairs of Conversational Breakdowns in
  Task-Oriented Dialogs}. In \bibinfo{booktitle}{\emph{Proceedings of the 33rd
  Annual ACM Symposium on User Interface Software and Technology}} (Virtual
  Event, USA) \emph{(\bibinfo{series}{UIST '20})}.
  \bibinfo{publisher}{Association for Computing Machinery},
  \bibinfo{address}{New York, NY, USA}, \bibinfo{pages}{1094–1107}.
\newblock
\showISBNx{9781450375146}
\urldef\tempurl%
\url{https://doi.org/10.1145/3379337.3415820}
\showDOI{\tempurl}


\bibitem[\protect\citeauthoryear{Liu, Chen, Lin, Chen, Irianti, Jen, Yeh, and
  Chiu}{Liu et~al\mbox{.}}{2020}]%
        {Liu2020}
\bibfield{author}{\bibinfo{person}{Ying-Chieh Liu}, \bibinfo{person}{Chien-Hung
  Chen}, \bibinfo{person}{Yu-Sheng Lin}, \bibinfo{person}{Hsin-Yun Chen},
  \bibinfo{person}{Denisa Irianti}, \bibinfo{person}{Ting-Ni Jen},
  \bibinfo{person}{Jou-Yin Yeh}, {and} \bibinfo{person}{Sherry Yueh-Hsia
  Chiu}.} \bibinfo{year}{2020}\natexlab{}.
\newblock \showarticletitle{Design and Usability Evaluation of Mobile
  Voice-Added Food Reporting for Elderly People: Randomized Controlled Trial}.
\newblock \bibinfo{journal}{\emph{JMIR Mhealth Uhealth}} \bibinfo{volume}{8},
  \bibinfo{number}{9} (\bibinfo{date}{28 Sep} \bibinfo{year}{2020}),
  \bibinfo{pages}{e20317}.
\newblock
\showISSN{2291-5222}
\urldef\tempurl%
\url{https://doi.org/10.2196/20317}
\showDOI{\tempurl}


\bibitem[\protect\citeauthoryear{Ma, Skubic, Ai, and Hubbard}{Ma
  et~al\mbox{.}}{2017}]%
        {Ma2017}
\bibfield{author}{\bibinfo{person}{Mengxuan Ma}, \bibinfo{person}{Majorie
  Skubic}, \bibinfo{person}{Karen Ai}, {and} \bibinfo{person}{Jordan Hubbard}.}
  \bibinfo{year}{2017}\natexlab{}.
\newblock \showarticletitle{Angel-Echo: A Personalized Health Care
  Application}. In \bibinfo{booktitle}{\emph{2017 IEEE/ACM International
  Conference on Connected Health: Applications, Systems and Engineering
  Technologies (CHASE)}} (Philadelphia, Pennsylvania, USA).
  \bibinfo{publisher}{Institute of Electrical and Electronics Engineers},
  \bibinfo{pages}{258 -- 259}.
\newblock
\showISBNx{9781509047222}
\urldef\tempurl%
\url{https://doi.org/10.1109/CHASE.2017.91}
\showDOI{\tempurl}


\bibitem[\protect\citeauthoryear{Marighetto, Etchamendy, Touzani, Torrea, Yee,
  Rawlins, and Jaffard}{Marighetto et~al\mbox{.}}{2001}]%
        {Marighetto2001}
\bibfield{author}{\bibinfo{person}{Aline Marighetto}, \bibinfo{person}{Nicole
  Etchamendy}, \bibinfo{person}{Khalid Touzani}, \bibinfo{person}{Cedric
  Torrea}, \bibinfo{person}{Benjamin Yee}, \bibinfo{person}{John Rawlins},
  {and} \bibinfo{person}{Robert Jaffard}.} \bibinfo{year}{2001}\natexlab{}.
\newblock \showarticletitle{Knowing {W}hich and {K}nowing {W}hat: A {P}otential
  {M}ouse {M}odel for {A}ge-related {H}uman {D}eclarative {M}emory {D}ecline}.
\newblock \bibinfo{journal}{\emph{European Journal of Neuroscience}}
  \bibinfo{volume}{11} (\bibinfo{date}{12} \bibinfo{year}{2001}),
  \bibinfo{pages}{3312 -- 3322}.
\newblock
\urldef\tempurl%
\url{https://doi.org/10.1046/j.1460-9568.1999.00741.x}
\showDOI{\tempurl}


\bibitem[\protect\citeauthoryear{McDonald, Schoenebeck, and Forte}{McDonald
  et~al\mbox{.}}{2019}]%
        {Nora2019}
\bibfield{author}{\bibinfo{person}{Nora McDonald}, \bibinfo{person}{Sarita
  Schoenebeck}, {and} \bibinfo{person}{Andrea Forte}.}
  \bibinfo{year}{2019}\natexlab{}.
\newblock \showarticletitle{Reliability and Inter-Rater Reliability in
  Qualitative Research: Norms and Guidelines for CSCW and HCI Practice}.
\newblock \bibinfo{journal}{\emph{Proc. ACM Hum.-Comput. Interact.}}
  \bibinfo{volume}{3}, \bibinfo{number}{CSCW}, Article \bibinfo{articleno}{72}
  (\bibinfo{date}{Nov.} \bibinfo{year}{2019}), \bibinfo{numpages}{23}~pages.
\newblock
\urldef\tempurl%
\url{https://doi.org/10.1145/3359174}
\showDOI{\tempurl}


\bibitem[\protect\citeauthoryear{Mercer, Giangregorio, Schneider, Chilana, Li1,
  and Grindrod}{Mercer et~al\mbox{.}}{2016}]%
        {Mercer2016}
\bibfield{author}{\bibinfo{person}{Kathryn Mercer}, \bibinfo{person}{Lora
  Giangregorio}, \bibinfo{person}{Eric Schneider}, \bibinfo{person}{Parmit
  Chilana}, \bibinfo{person}{Melissa Li1}, {and} \bibinfo{person}{Kelly
  Grindrod}.} \bibinfo{year}{2016}\natexlab{}.
\newblock \showarticletitle{Acceptance of Commercially Available Wearable
  Activity Trackers Among Adults Aged Over 50 and With Chronic Illness: A
  Mixed-Methods Evaluation}.
\newblock \bibinfo{journal}{\emph{Journal of Medical Internet Research}}
  \bibinfo{volume}{4}, \bibinfo{number}{1} (\bibinfo{year}{2016}).
\newblock
\showISSN{1438-8871}
\urldef\tempurl%
\url{https://doi.org/10.2196/mhealth.4225}
\showDOI{\tempurl}


\bibitem[\protect\citeauthoryear{Mitzner, Boron, Fausset, Adams, Charness,
  Czaja, Dijkstra, Fisk, Rogers, and Sharit}{Mitzner et~al\mbox{.}}{2010}]%
        {Tracy2010}
\bibfield{author}{\bibinfo{person}{Tracy~L. Mitzner}, \bibinfo{person}{Julie~B.
  Boron}, \bibinfo{person}{Cara~Bailey Fausset}, \bibinfo{person}{Anne~E.
  Adams}, \bibinfo{person}{Neil Charness}, \bibinfo{person}{Sara~J. Czaja},
  \bibinfo{person}{Katinka Dijkstra}, \bibinfo{person}{Arthur~D. Fisk},
  \bibinfo{person}{Wendy~A. Rogers}, {and} \bibinfo{person}{Joseph Sharit}.}
  \bibinfo{year}{2010}\natexlab{}.
\newblock \showarticletitle{Older Adults Talk Technology: Technology Usage and
  Attitudes}.
\newblock \bibinfo{journal}{\emph{Computers in Human Behavior}}
  \bibinfo{volume}{26}, \bibinfo{number}{6} (\bibinfo{year}{2010}),
  \bibinfo{pages}{1710 -- 1721}.
\newblock
\showISSN{0747-5632}
\urldef\tempurl%
\url{https://doi.org/10.1016/j.chb.2010.06.020}
\showDOI{\tempurl}
\newblock
\shownote{Online Interactivity: Role of Technology in Behavior Change.}


\bibitem[\protect\citeauthoryear{Mrini, Chen, Nakashole, Weibel, and
  Farcas}{Mrini et~al\mbox{.}}{2021}]%
        {Mrini2021}
\bibfield{author}{\bibinfo{person}{Khalil Mrini}, \bibinfo{person}{Chen Chen},
  \bibinfo{person}{Ndapa Nakashole}, \bibinfo{person}{Nadir Weibel}, {and}
  \bibinfo{person}{Emilia Farcas}.} \bibinfo{year}{2021}\natexlab{}.
\newblock \showarticletitle{Medical Question Understanding and Answering for
  Older Adults}.
\newblock \bibinfo{journal}{\emph{Southern California Machine Learning and
  Natural Language Processing Symposium}} (\bibinfo{year}{2021}).
\newblock


\bibitem[\protect\citeauthoryear{Mundt}{Mundt}{1997}]%
        {James1997}
\bibfield{author}{\bibinfo{person}{James~C Mundt}.}
  \bibinfo{year}{1997}\natexlab{}.
\newblock \showarticletitle{Interactive Voice Response Systems in Clinical
  Research and Treatment}.
\newblock \bibinfo{journal}{\emph{Psychiatr Serv}} \bibinfo{volume}{48},
  \bibinfo{number}{5} (\bibinfo{year}{1997}), \bibinfo{pages}{611--612}.
\newblock
\urldef\tempurl%
\url{https://doi.org/10.4103/2229-3485.173781}
\showDOI{\tempurl}


\bibitem[\protect\citeauthoryear{Mynatt and Rogers}{Mynatt and Rogers}{2 01}]%
        {Elizabeth2001}
\bibfield{author}{\bibinfo{person}{Elizabeth~D. Mynatt} {and}
  \bibinfo{person}{Wendy~A. Rogers}.} \bibinfo{year}{2001-12-01}\natexlab{}.
\newblock \showarticletitle{Developing Technology to Support the Functional
  Independence of Older Adults}.
\newblock \bibinfo{journal}{\emph{Ageing International}} \bibinfo{volume}{27},
  \bibinfo{number}{1} (\bibinfo{year}{2001-12-01}), \bibinfo{pages}{24 -- 41}.
\newblock
\showISSN{1936-606X}
\urldef\tempurl%
\url{https://doi.org/10.1007/s12126-001-1014-5}
\showDOI{\tempurl}


\bibitem[\protect\citeauthoryear{Nations}{Nations}{2020}]%
        {un_population_aging}
\bibfield{author}{\bibinfo{person}{United Nations}.}
  \bibinfo{year}{2020}\natexlab{}.
\newblock \showarticletitle{World Population Ageing 2019 -- Highlights}.
\newblock  (\bibinfo{year}{2020}).
\newblock
\urldef\tempurl%
\url{https://www.un.org/en/development/desa/population/publications/pdf/ageing/WorldPopulationAgeing2019-Highlights.pdf}
\showURL{%
\tempurl}


\bibitem[\protect\citeauthoryear{Newall and Menec}{Newall and Menec}{2019}]%
        {Newall2019}
\bibfield{author}{\bibinfo{person}{Nancy E.~G. Newall} {and}
  \bibinfo{person}{Verena~H. Menec}.} \bibinfo{year}{2019}\natexlab{}.
\newblock \showarticletitle{{L}oneliness and {S}ocial {I}solation of {O}lder
  {A}dults: {W}hy {I}t {I}s {I}mportant to {E}xamine these {S}ocial {A}spects
  {T}ogether}.
\newblock \bibinfo{journal}{\emph{Journal of Social and Personal
  Relationships}} \bibinfo{volume}{36}, \bibinfo{number}{3}
  (\bibinfo{year}{2019}), \bibinfo{pages}{925--939}.
\newblock
\urldef\tempurl%
\url{https://doi.org/10.1177/0265407517749045}
\showDOI{\tempurl}


\bibitem[\protect\citeauthoryear{Niazkhani, Toni, Cheshmekaboodi, Georgiou, and
  Pirnejad}{Niazkhani et~al\mbox{.}}{2020}]%
        {Niazkhani2020}
\bibfield{author}{\bibinfo{person}{Zahra Niazkhani}, \bibinfo{person}{Esmaeel
  Toni}, \bibinfo{person}{Mojgan Cheshmekaboodi}, \bibinfo{person}{Andrew
  Georgiou}, {and} \bibinfo{person}{Habibollah Pirnejad}.}
  \bibinfo{year}{2020}\natexlab{}.
\newblock \showarticletitle{Barriers to Patient, Provider, and Caregiver
  Adoption and Use of Electronic Personal Health Records in Chronic Care: a
  Systematic Review}.
\newblock \bibinfo{journal}{\emph{BMC Medical Informatics and Decision Making}}
  (\bibinfo{date}{Jan} \bibinfo{year}{2020}).
\newblock
\urldef\tempurl%
\url{https://doi.org/10.21203/rs.2.22158/v1}
\showDOI{\tempurl}


\bibitem[\protect\citeauthoryear{O{\textquoteright}Brien, Liggett,
  Ramirez-Zohfeld, Sunkara, and Lindquist}{O{\textquoteright}Brien
  et~al\mbox{.}}{2020}]%
        {Katherine2020}
\bibfield{author}{\bibinfo{person}{Katherine O{\textquoteright}Brien},
  \bibinfo{person}{Anna Liggett}, \bibinfo{person}{Vanessa Ramirez-Zohfeld},
  \bibinfo{person}{Priya Sunkara}, {and} \bibinfo{person}{Lee~A Lindquist}.}
  \bibinfo{year}{2020}\natexlab{}.
\newblock \bibinfo{title}{Voice-Controlled Intelligent Personal Assistants to
  Support Aging in Place}.
\newblock , \bibinfo{numpages}{176 -- 179}~pages.
\newblock
\urldef\tempurl%
\url{https://doi.org/10.1111/jgs.16217}
\showDOI{\tempurl}


\bibitem[\protect\citeauthoryear{Pang, Collin~Wang, McGrenere, Leung, Dai, and
  Moffatt}{Pang et~al\mbox{.}}{2021}]%
        {Pang2021}
\bibfield{author}{\bibinfo{person}{Carolyn Pang}, \bibinfo{person}{Zhiqin
  Collin~Wang}, \bibinfo{person}{Joanna McGrenere}, \bibinfo{person}{Rock
  Leung}, \bibinfo{person}{Jiamin Dai}, {and} \bibinfo{person}{Karyn Moffatt}.}
  \bibinfo{year}{2021}\natexlab{}.
\newblock \bibinfo{booktitle}{\emph{Technology Adoption and Learning
  Preferences for Older Adults: Evolving Perceptions, Ongoing Challenges, and
  Emerging Design Opportunities}}.
\newblock \bibinfo{publisher}{Association for Computing Machinery},
  \bibinfo{address}{New York, NY, USA}.
\newblock
\showISBNx{9781450380966}
\urldef\tempurl%
\url{https://doi.org/10.1145/3411764.3445702}
\showURL{%
\tempurl}


\bibitem[\protect\citeauthoryear{Park}{Park}{1998}]%
        {Park1998}
\bibfield{author}{\bibinfo{person}{Denise~C Park}.}
  \bibinfo{year}{1998}\natexlab{}.
\newblock \showarticletitle{{A}geing and {M}emory: {M}echanisms {U}nderlying
  {A}ge {D}ifferences in {P}erformance}.
\newblock \bibinfo{journal}{\emph{Australasian Journal on Ageing}}
  \bibinfo{volume}{17} (\bibinfo{year}{1998}), \bibinfo{pages}{69--72}.
\newblock


\bibitem[\protect\citeauthoryear{Piper, Campbell, and Hollan}{Piper
  et~al\mbox{.}}{2010}]%
        {Piper2010}
\bibfield{author}{\bibinfo{person}{Anne~Marie Piper}, \bibinfo{person}{Ross
  Campbell}, {and} \bibinfo{person}{James~D. Hollan}.}
  \bibinfo{year}{2010}\natexlab{}.
\newblock \showarticletitle{Exploring the Accessibility and Appeal of Surface
  Computing for Older Adult Health Care Support}. In
  \bibinfo{booktitle}{\emph{Proceedings of the SIGCHI Conference on Human
  Factors in Computing Systems}} (Atlanta, Georgia, USA)
  \emph{(\bibinfo{series}{CHI '10})}. \bibinfo{publisher}{Association for
  Computing Machinery}, \bibinfo{address}{New York, NY, USA},
  \bibinfo{pages}{907–916}.
\newblock
\showISBNx{9781605589299}
\urldef\tempurl%
\url{https://doi.org/10.1145/1753326.1753461}
\showDOI{\tempurl}


\bibitem[\protect\citeauthoryear{Piper, Cornejo, and Brewer}{Piper
  et~al\mbox{.}}{2016}]%
        {Piper2016}
\bibfield{author}{\bibinfo{person}{Anne~Marie Piper}, \bibinfo{person}{Raymundo
  Cornejo}, {and} \bibinfo{person}{Robin Brewer}.}
  \bibinfo{year}{2016}\natexlab{}.
\newblock \showarticletitle{Understanding the Challenges and Opportunities of
  Smart Mobile Devices among the Oldest Old}.
\newblock \bibinfo{journal}{\emph{International Journal of Mobile Human
  Computer Interaction}}  \bibinfo{volume}{8} (\bibinfo{date}{04}
  \bibinfo{year}{2016}), \bibinfo{pages}{83--98}.
\newblock
\urldef\tempurl%
\url{https://doi.org/10.4018/IJMHCI.2016040105}
\showDOI{\tempurl}


\bibitem[\protect\citeauthoryear{Porcheron, Fischer, Reeves, and
  Sharples}{Porcheron et~al\mbox{.}}{2018}]%
        {Porcheron2018}
\bibfield{author}{\bibinfo{person}{Martin Porcheron}, \bibinfo{person}{Joel~E.
  Fischer}, \bibinfo{person}{Stuart Reeves}, {and} \bibinfo{person}{Sarah
  Sharples}.} \bibinfo{year}{2018}\natexlab{}.
\newblock \showarticletitle{Voice Interfaces in Everyday Life}. In
  \bibinfo{booktitle}{\emph{Proceedings of the 2018 CHI Conference on Human
  Factors in Computing Systems}} (Montreal QC, Canada)
  \emph{(\bibinfo{series}{CHI '18})}. \bibinfo{publisher}{Association for
  Computing Machinery}, \bibinfo{address}{New York, NY, USA},
  \bibinfo{pages}{1–12}.
\newblock
\showISBNx{9781450356206}
\urldef\tempurl%
\url{https://doi.org/10.1145/3173574.3174214}
\showDOI{\tempurl}


\bibitem[\protect\citeauthoryear{Pradhan, Findlater, and Lazar}{Pradhan
  et~al\mbox{.}}{2019}]%
        {Alisha2019}
\bibfield{author}{\bibinfo{person}{Alisha Pradhan}, \bibinfo{person}{Leah
  Findlater}, {and} \bibinfo{person}{Amanda Lazar}.}
  \bibinfo{year}{2019}\natexlab{}.
\newblock \showarticletitle{“Phantom Friend” or “Just a Box with
  Information”: Personification and Ontological Categorization of Smart
  Speaker-Based Voice Assistants by Older Adults}.
\newblock \bibinfo{journal}{\emph{Proc. ACM Hum.-Comput. Interact.}}
  \bibinfo{volume}{3}, \bibinfo{number}{CSCW}, Article \bibinfo{articleno}{214}
  (\bibinfo{date}{Nov.} \bibinfo{year}{2019}), \bibinfo{numpages}{21}~pages.
\newblock
\urldef\tempurl%
\url{https://doi.org/10.1145/3359316}
\showDOI{\tempurl}


\bibitem[\protect\citeauthoryear{Pradhan, Lazar, and Findlater}{Pradhan
  et~al\mbox{.}}{2020}]%
        {Pradhan2020}
\bibfield{author}{\bibinfo{person}{Alisha Pradhan}, \bibinfo{person}{Amanda
  Lazar}, {and} \bibinfo{person}{Leah Findlater}.}
  \bibinfo{year}{2020}\natexlab{}.
\newblock \showarticletitle{Use of Intelligent Voice Assistants by Older Adults
  with Low Technology Use}.
\newblock \bibinfo{journal}{\emph{ACM Trans. Comput.-Hum. Interact.}}
  \bibinfo{volume}{27}, \bibinfo{number}{4}, Article \bibinfo{articleno}{31}
  (\bibinfo{date}{Sept.} \bibinfo{year}{2020}), \bibinfo{numpages}{27}~pages.
\newblock
\showISSN{1073-0516}
\urldef\tempurl%
\url{https://doi.org/10.1145/3373759}
\showDOI{\tempurl}


\bibitem[\protect\citeauthoryear{Ramirez-Zohfeld, Seltzer, Xiong, Morse, and
  Lindquist}{Ramirez-Zohfeld et~al\mbox{.}}{2020}]%
        {Ramirez2020}
\bibfield{author}{\bibinfo{person}{Vanessa Ramirez-Zohfeld},
  \bibinfo{person}{Anne Seltzer}, \bibinfo{person}{Linda Xiong},
  \bibinfo{person}{Lucy Morse}, {and} \bibinfo{person}{Lee~A. Lindquist}.}
  \bibinfo{year}{2020}\natexlab{}.
\newblock \showarticletitle{{U}se of {E}lectronic {H}ealth {R}ecords by {O}lder
  {A}dults, 85 {Y}ears and {O}lder, and {T}heir {C}aregivers}.
\newblock \bibinfo{journal}{\emph{J Am Geriatr Soc}} \bibinfo{volume}{68},
  \bibinfo{number}{5} (\bibinfo{date}{May} \bibinfo{year}{2020}),
  \bibinfo{pages}{1078--1082}.
\newblock
\urldef\tempurl%
\url{https://doi.org/10.1111/jgs.16393}
\showDOI{\tempurl}


\bibitem[\protect\citeauthoryear{Ruan, Wobbrock, Liou, Ng, and Landay}{Ruan
  et~al\mbox{.}}{2016}]%
        {Ruan2016}
\bibfield{author}{\bibinfo{person}{Sherry Ruan}, \bibinfo{person}{Jacob
  Wobbrock}, \bibinfo{person}{Kenny Liou}, \bibinfo{person}{Andrew Ng}, {and}
  \bibinfo{person}{James Landay}.} \bibinfo{year}{2016}\natexlab{}.
\newblock \showarticletitle{Speech Is 3x Faster than Typing for English and
  Mandarin Text Entry on Mobile Devices}.
\newblock  (\bibinfo{date}{08} \bibinfo{year}{2016}).
\newblock
\urldef\tempurl%
\url{https://arxiv.org/pdf/1608.07323v1.pdf}
\showURL{%
\tempurl}


\bibitem[\protect\citeauthoryear{Sanders and Stappers}{Sanders and
  Stappers}{2008}]%
        {Sanders2008}
\bibfield{author}{\bibinfo{person}{Elizabeth B.-N. Sanders} {and}
  \bibinfo{person}{Pieter~Jan Stappers}.} \bibinfo{year}{2008}\natexlab{}.
\newblock \showarticletitle{Co-creation and the new landscapes of design}.
\newblock \bibinfo{journal}{\emph{CoDesign}} \bibinfo{volume}{4},
  \bibinfo{number}{1} (\bibinfo{year}{2008}), \bibinfo{pages}{5--18}.
\newblock
\urldef\tempurl%
\url{https://doi.org/10.1080/15710880701875068}
\showDOI{\tempurl}


\bibitem[\protect\citeauthoryear{Sanders and Martin-Hammond}{Sanders and
  Martin-Hammond}{2019}]%
        {Jamie2019}
\bibfield{author}{\bibinfo{person}{Jamie Sanders} {and}
  \bibinfo{person}{Aqueasha Martin-Hammond}.} \bibinfo{year}{2019}\natexlab{}.
\newblock \showarticletitle{Exploring Autonomy in the Design of an Intelligent
  Health Assistant for Older Adults}. In \bibinfo{booktitle}{\emph{Proceedings
  of the 24th International Conference on Intelligent User Interfaces:
  Companion}} (Marina del Ray, California) \emph{(\bibinfo{series}{IUI '19})}.
  \bibinfo{publisher}{Association for Computing Machinery},
  \bibinfo{address}{New York, NY, USA}, \bibinfo{pages}{95–96}.
\newblock
\showISBNx{9781450366731}
\urldef\tempurl%
\url{https://doi.org/10.1145/3308557.3308713}
\showDOI{\tempurl}


\bibitem[\protect\citeauthoryear{Sato, Kobayashi, Takagi, Asakawa, and
  Tanaka}{Sato et~al\mbox{.}}{2011}]%
        {Sato2011}
\bibfield{author}{\bibinfo{person}{Daisuke Sato}, \bibinfo{person}{Masatomo
  Kobayashi}, \bibinfo{person}{Hironobu Takagi}, \bibinfo{person}{Chieko
  Asakawa}, {and} \bibinfo{person}{Jiro Tanaka}.}
  \bibinfo{year}{2011}\natexlab{}.
\newblock \showarticletitle{How Voice Augmentation Supports Elderly Web Users}.
  In \bibinfo{booktitle}{\emph{The Proceedings of the 13th International ACM
  SIGACCESS Conference on Computers and Accessibility}} (Dundee, Scotland, UK)
  \emph{(\bibinfo{series}{ASSETS '11})}. \bibinfo{publisher}{Association for
  Computing Machinery}, \bibinfo{address}{New York, NY, USA},
  \bibinfo{pages}{155–162}.
\newblock
\showISBNx{9781450309202}
\urldef\tempurl%
\url{https://doi.org/10.1145/2049536.2049565}
\showDOI{\tempurl}


\bibitem[\protect\citeauthoryear{Sayago, Neves, and Cowan}{Sayago
  et~al\mbox{.}}{2019}]%
        {Sergio2019}
\bibfield{author}{\bibinfo{person}{Sergio Sayago},
  \bibinfo{person}{Barbara~Barbosa Neves}, {and} \bibinfo{person}{Benjamin~R
  Cowan}.} \bibinfo{year}{2019}\natexlab{}.
\newblock \showarticletitle{Voice Assistants and Older People: Some Open
  Issues}. In \bibinfo{booktitle}{\emph{Proceedings of the 1st International
  Conference on Conversational User Interfaces}} (Dublin, Ireland)
  \emph{(\bibinfo{series}{CUI ’19})}. \bibinfo{publisher}{Association for
  Computing Machinery}, \bibinfo{address}{New York, NY, USA}, Article
  \bibinfo{articleno}{7}, \bibinfo{numpages}{3}~pages.
\newblock
\showISBNx{9781450371872}
\urldef\tempurl%
\url{https://doi.org/10.1145/3342775.3342803}
\showDOI{\tempurl}


\bibitem[\protect\citeauthoryear{Schl\"{o}gl, Chollet, Garschall, Tscheligi,
  and Legouverneur}{Schl\"{o}gl et~al\mbox{.}}{2013}]%
        {Schlogl2013}
\bibfield{author}{\bibinfo{person}{S. Schl\"{o}gl}, \bibinfo{person}{G.
  Chollet}, \bibinfo{person}{M. Garschall}, \bibinfo{person}{M. Tscheligi},
  {and} \bibinfo{person}{G. Legouverneur}.} \bibinfo{year}{2013}\natexlab{}.
\newblock \showarticletitle{Exploring Voice User Interfaces for Seniors}. In
  \bibinfo{booktitle}{\emph{Proceedings of the 6th International Conference on
  Pervasive Technologies Related to Assistive Environments}} (Rhodes, Greece)
  \emph{(\bibinfo{series}{PETRA ’13})}. \bibinfo{publisher}{Association for
  Computing Machinery}, \bibinfo{address}{New York, NY, USA}, Article
  \bibinfo{articleno}{52}, \bibinfo{numpages}{2}~pages.
\newblock
\showISBNx{9781450319737}
\urldef\tempurl%
\url{https://doi.org/10.1145/2504335.2504391}
\showDOI{\tempurl}


\bibitem[\protect\citeauthoryear{Smith and Chaparro}{Smith and
  Chaparro}{2015}]%
        {Smith2015}
\bibfield{author}{\bibinfo{person}{Amanda~L. Smith} {and}
  \bibinfo{person}{Barbara~S. Chaparro}.} \bibinfo{year}{2015}\natexlab{}.
\newblock \showarticletitle{Smartphone Text Input Method Performance,
  Usability, and Preference With Younger and Older Adults}.
\newblock \bibinfo{journal}{\emph{Human Factors}} \bibinfo{volume}{57},
  \bibinfo{number}{6} (\bibinfo{year}{2015}), \bibinfo{pages}{1015--1028}.
\newblock
\urldef\tempurl%
\url{https://doi.org/10.1177/0018720815575644}
\showDOI{\tempurl}


\bibitem[\protect\citeauthoryear{Statista}{Statista}{2020}]%
        {SDGDP}
\bibfield{author}{\bibinfo{person}{Statista}.} \bibinfo{year}{2020}\natexlab{}.
\newblock \bibinfo{booktitle}{\emph{Forecasted Gross Metropolitan Product (GMP)
  of the United States in 2020, by metropolitan area}}.
\newblock
\urldef\tempurl%
\url{https://www.statista.com/statistics/183808/gmp-of-the-20-biggest-metro-areas/}
\showURL{%
\tempurl}


\bibitem[\protect\citeauthoryear{Strauss and Corbin}{Strauss and
  Corbin}{1994}]%
        {strauss1994grounded}
\bibfield{author}{\bibinfo{person}{Anselm Strauss} {and}
  \bibinfo{person}{Juliet Corbin}.} \bibinfo{year}{1994}\natexlab{}.
\newblock \showarticletitle{Grounded Theory Methodology}.
\newblock \bibinfo{journal}{\emph{Handbook of Qualitative Research}}
  \bibinfo{volume}{17}, \bibinfo{number}{1} (\bibinfo{year}{1994}),
  \bibinfo{pages}{273--285}.
\newblock


\bibitem[\protect\citeauthoryear{Sun, Chen, and Zhang}{Sun
  et~al\mbox{.}}{2020}]%
        {Ke2020}
\bibfield{author}{\bibinfo{person}{Ke Sun}, \bibinfo{person}{Chen Chen}, {and}
  \bibinfo{person}{Xinyu Zhang}.} \bibinfo{year}{2020}\natexlab{}.
\newblock \showarticletitle{"Alexa, Stop Spying on Me!": Speech Privacy
  Protection against Voice Assistants}. In
  \bibinfo{booktitle}{\emph{Proceedings of the 18th Conference on Embedded
  Networked Sensor Systems}}. \bibinfo{publisher}{Association for Computing
  Machinery}, \bibinfo{address}{New York, NY, USA}, \bibinfo{pages}{298–311}.
\newblock
\showISBNx{9781450375900}
\urldef\tempurl%
\url{https://doi.org/10.1145/3384419.3430727}
\showURL{%
\tempurl}


\bibitem[\protect\citeauthoryear{Takagi, Ohno, Kobayashi, and Nakada}{Takagi
  et~al\mbox{.}}{2018}]%
        {Takagi2018}
\bibfield{author}{\bibinfo{person}{Hironobu Takagi}, \bibinfo{person}{Masaki
  Ohno}, \bibinfo{person}{Masatomo Kobayashi}, {and} \bibinfo{person}{Takeo
  Nakada}.} \bibinfo{year}{2018}\natexlab{}.
\newblock \showarticletitle{{E}valuating {S}peech -- {B}ased {Q}uestion --
  {A}nswer {I}nteractions for {E}lder -- {C}are {S}ervices}.
\newblock \bibinfo{journal}{\emph{IBM Journal of Research and Development}}
  \bibinfo{volume}{62}, \bibinfo{number}{1} (\bibinfo{year}{2018}),
  \bibinfo{pages}{6:1--6:10}.
\newblock
\urldef\tempurl%
\url{https://doi.org/10.1147/JRD.2017.2768720}
\showDOI{\tempurl}


\bibitem[\protect\citeauthoryear{Teixeira, Ferreira, Almeida, Silva, Rosa,
  Pereira, and Vieira}{Teixeira et~al\mbox{.}}{2016}]%
        {Antonio2017}
\bibfield{author}{\bibinfo{person}{Ant{\'{o}}nio Teixeira},
  \bibinfo{person}{Fl{\'{a}}vio Ferreira}, \bibinfo{person}{Nuno Almeida},
  \bibinfo{person}{Samuel Silva}, \bibinfo{person}{Ana~Filipa Rosa},
  \bibinfo{person}{Jos{\'{e}}~Casimiro Pereira}, {and} \bibinfo{person}{Diogo
  Vieira}.} \bibinfo{year}{2016}\natexlab{}.
\newblock \showarticletitle{Design and Development of Medication Assistant:
  Older Adults Centred Design to go beyond Simple Medication Reminders}.
\newblock \bibinfo{journal}{\emph{Universal Access in the Information Society}}
  \bibinfo{volume}{16}, \bibinfo{number}{3} (\bibinfo{date}{Jul}
  \bibinfo{year}{2016}), \bibinfo{pages}{545 -- 560}.
\newblock
\showISSN{1615-5297}
\urldef\tempurl%
\url{https://doi.org/10.1007/s10209-016-0487-7}
\showDOI{\tempurl}


\bibitem[\protect\citeauthoryear{Trajkova and Martin-Hammond}{Trajkova and
  Martin-Hammond}{2020}]%
        {Milka2020}
\bibfield{author}{\bibinfo{person}{Milka Trajkova} {and}
  \bibinfo{person}{Aqueasha Martin-Hammond}.} \bibinfo{year}{2020}\natexlab{}.
\newblock \showarticletitle{``Alexa is a Toy'': Exploring Older Adults’
  Reasons for Using, Limiting, and Abandoning Echo}. In
  \bibinfo{booktitle}{\emph{Proceedings of the 2020 CHI Conference on Human
  Factors in Computing Systems}} (Honolulu, HI, USA)
  \emph{(\bibinfo{series}{CHI ’20})}. \bibinfo{publisher}{Association for
  Computing Machinery}, \bibinfo{address}{New York, NY, USA},
  \bibinfo{pages}{1–13}.
\newblock
\showISBNx{9781450367080}
\urldef\tempurl%
\url{https://doi.org/10.1145/3313831.3376760}
\showDOI{\tempurl}


\bibitem[\protect\citeauthoryear{Vines, Pritchard, Wright, Olivier, and
  Brittain}{Vines et~al\mbox{.}}{2015}]%
        {Vines2015}
\bibfield{author}{\bibinfo{person}{John Vines}, \bibinfo{person}{Gary
  Pritchard}, \bibinfo{person}{Peter Wright}, \bibinfo{person}{Patrick
  Olivier}, {and} \bibinfo{person}{Katie Brittain}.}
  \bibinfo{year}{2015}\natexlab{}.
\newblock \showarticletitle{An Age-Old Problem: Examining the Discourses of
  Ageing in HCI and Strategies for Future Research}.
\newblock \bibinfo{journal}{\emph{ACM Trans. Comput.-Hum. Interact.}}
  \bibinfo{volume}{22}, \bibinfo{number}{1}, Article \bibinfo{articleno}{2}
  (\bibinfo{date}{Feb.} \bibinfo{year}{2015}), \bibinfo{numpages}{27}~pages.
\newblock
\showISSN{1073-0516}
\urldef\tempurl%
\url{https://doi.org/10.1145/2696867}
\showDOI{\tempurl}


\bibitem[\protect\citeauthoryear{Wang, Carroll, Peck, Myneni, and Gong}{Wang
  et~al\mbox{.}}{2016}]%
        {Wang2016}
\bibfield{author}{\bibinfo{person}{Jing Wang}, \bibinfo{person}{Deidra
  Carroll}, \bibinfo{person}{Michelle Peck}, \bibinfo{person}{Sahiti Myneni},
  {and} \bibinfo{person}{Yang Gong}.} \bibinfo{year}{2016}\natexlab{}.
\newblock \showarticletitle{Mobile and Wearable Technology Needs for Aging in
  Place: Perspectives from Older Adults and Their Caregivers and Providers}.
\newblock \bibinfo{journal}{\emph{Stud Health Technol Inform}}
  (\bibinfo{year}{2016}).
\newblock
\urldef\tempurl%
\url{https://pubmed.ncbi.nlm.nih.gov/27332248/}
\showURL{%
\tempurl}


\bibitem[\protect\citeauthoryear{Weiner, Callahan, Tierney, Overhage, Mamlin,
  Dexter, and McDonald}{Weiner et~al\mbox{.}}{2003}]%
        {Weiner2003}
\bibfield{author}{\bibinfo{person}{Michael Weiner},
  \bibinfo{person}{Christopher~M. Callahan}, \bibinfo{person}{William~M.
  Tierney}, \bibinfo{person}{J.~Marc Overhage}, \bibinfo{person}{Burke Mamlin},
  \bibinfo{person}{Paul~R. Dexter}, {and} \bibinfo{person}{Clement~J.
  McDonald}.} \bibinfo{year}{2003}\natexlab{}.
\newblock \showarticletitle{Using Information Technology To Improve the Health
  Care of Older Adults}.
\newblock \bibinfo{journal}{\emph{Annals of Internal Medicine}}
  \bibinfo{volume}{139}, \bibinfo{number}{5\_Part\_2} (\bibinfo{year}{2003}),
  \bibinfo{pages}{430--436}.
\newblock
\urldef\tempurl%
\url{https://doi.org/10.7326/0003-4819-139-5_Part_2-200309021-00010}
\showDOI{\tempurl}


\bibitem[\protect\citeauthoryear{Whiting}{Whiting}{2008}]%
        {Lisa2008}
\bibfield{author}{\bibinfo{person}{Lisa~S. Whiting}.}
  \bibinfo{year}{2008}\natexlab{}.
\newblock \showarticletitle{Semi-Structured Interviews: Guidance for Novice
  Researchers}.
\newblock \bibinfo{journal}{\emph{Nursing Standard}} \bibinfo{volume}{22},
  \bibinfo{number}{23} (\bibinfo{year}{2008}).
\newblock
\showISSN{00296570}
\urldef\tempurl%
\url{https://doi.org/10.7748/ns2008.02.22.23.35.c6420}
\showDOI{\tempurl}


\bibitem[\protect\citeauthoryear{Wobbrock, Gajos, Kane, and
  Vanderheiden}{Wobbrock et~al\mbox{.}}{2018}]%
        {Wobbrock2018}
\bibfield{author}{\bibinfo{person}{Jacob~O. Wobbrock},
  \bibinfo{person}{Krzysztof~Z. Gajos}, \bibinfo{person}{Shaun~K. Kane}, {and}
  \bibinfo{person}{Gregg~C. Vanderheiden}.} \bibinfo{year}{2018}\natexlab{}.
\newblock \showarticletitle{Ability-Based Design}.
\newblock \bibinfo{journal}{\emph{Commun. ACM}} \bibinfo{volume}{61},
  \bibinfo{number}{6} (\bibinfo{date}{May} \bibinfo{year}{2018}),
  \bibinfo{pages}{62–71}.
\newblock
\showISSN{0001-0782}
\urldef\tempurl%
\url{https://doi.org/10.1145/3148051}
\showDOI{\tempurl}


\bibitem[\protect\citeauthoryear{Wobbrock, Kane, Gajos, Harada, and
  Froehlich}{Wobbrock et~al\mbox{.}}{2011}]%
        {Wobbrock2011}
\bibfield{author}{\bibinfo{person}{Jacob~O. Wobbrock},
  \bibinfo{person}{Shaun~K. Kane}, \bibinfo{person}{Krzysztof~Z. Gajos},
  \bibinfo{person}{Susumu Harada}, {and} \bibinfo{person}{Jon Froehlich}.}
  \bibinfo{year}{2011}\natexlab{}.
\newblock \showarticletitle{Ability-Based Design: Concept, Principles and
  Examples}.
\newblock \bibinfo{journal}{\emph{ACM Transactions on Accessible Computing}}
  \bibinfo{volume}{3}, \bibinfo{number}{3}, Article \bibinfo{articleno}{9}
  (\bibinfo{date}{April} \bibinfo{year}{2011}), \bibinfo{numpages}{27}~pages.
\newblock
\showISSN{1936-7228}
\urldef\tempurl%
\url{https://doi.org/10.1145/1952383.1952384}
\showDOI{\tempurl}


\bibitem[\protect\citeauthoryear{Wulf, Garschall, Himmelsbach, and
  Tscheligi}{Wulf et~al\mbox{.}}{2014}]%
        {Wulf2014}
\bibfield{author}{\bibinfo{person}{Linda Wulf}, \bibinfo{person}{Markus
  Garschall}, \bibinfo{person}{Julia Himmelsbach}, {and}
  \bibinfo{person}{Manfred Tscheligi}.} \bibinfo{year}{2014}\natexlab{}.
\newblock \showarticletitle{Hands Free - Care Free: Elderly People Taking
  Advantage of Speech-Only Interaction}. In
  \bibinfo{booktitle}{\emph{Proceedings of the 8th Nordic Conference on
  Human-Computer Interaction: Fun, Fast, Foundational}} (Helsinki, Finland)
  \emph{(\bibinfo{series}{NordiCHI '14})}. \bibinfo{publisher}{Association for
  Computing Machinery}, \bibinfo{address}{New York, NY, USA},
  \bibinfo{pages}{203–206}.
\newblock
\showISBNx{9781450325424}
\urldef\tempurl%
\url{https://doi.org/10.1145/2639189.2639251}
\showDOI{\tempurl}


\bibitem[\protect\citeauthoryear{Yarmand, Chen, Gasques, Murphy, and
  Weibel}{Yarmand et~al\mbox{.}}{2021}]%
        {Yarmand2021}
\bibfield{author}{\bibinfo{person}{Matin Yarmand}, \bibinfo{person}{Chen Chen},
  \bibinfo{person}{Danilo Gasques}, \bibinfo{person}{James~D. Murphy}, {and}
  \bibinfo{person}{Nadir Weibel}.} \bibinfo{year}{2021}\natexlab{}.
\newblock \showarticletitle{Facilitating Remote Design Thinking Workshops in
  Healthcare: the Case of Contouring in Radiation Oncology}. In
  \bibinfo{booktitle}{\emph{Proceedings of the 2021 ACM Conference on Human
  Factors in Computing Systems}} (Yokohama, Japn) \emph{(\bibinfo{series}{CHI
  '21})}. \bibinfo{publisher}{Association for Computing Machinery},
  \bibinfo{address}{Yokohama, Japan}, 5.
\newblock
\urldef\tempurl%
\url{https://doi.org/10.1145/3411763.3443445}
\showDOI{\tempurl}


\bibitem[\protect\citeauthoryear{Zajicek}{Zajicek}{2001}]%
        {Zajicek2001}
\bibfield{author}{\bibinfo{person}{Mary Zajicek}.}
  \bibinfo{year}{2001}\natexlab{}.
\newblock \showarticletitle{Interface Design for Older Adults}. In
  \bibinfo{booktitle}{\emph{Proceedings of the 2001 EC/NSF Workshop on
  Universal Accessibility of Ubiquitous Computing: Providing for the Elderly}}
  (Alc\'{a}cer do Sal, Portugal) \emph{(\bibinfo{series}{WUAUC'01})}.
  \bibinfo{publisher}{Association for Computing Machinery},
  \bibinfo{address}{New York, NY, USA}, \bibinfo{pages}{60–65}.
\newblock
\showISBNx{158113424X}
\urldef\tempurl%
\url{https://doi.org/10.1145/564526.564543}
\showDOI{\tempurl}


\bibitem[\protect\citeauthoryear{Ziman and Walsh}{Ziman and Walsh}{2018}]%
        {Randall2018}
\bibfield{author}{\bibinfo{person}{Randall Ziman} {and} \bibinfo{person}{Greg
  Walsh}.} \bibinfo{year}{2018}\natexlab{}.
\newblock \showarticletitle{Factors Affecting Seniors' Perceptions of
  Voice-Enabled User Interfaces}. In \bibinfo{booktitle}{\emph{Extended
  Abstracts of the 2018 CHI Conference on Human Factors in Computing Systems}}
  \emph{(\bibinfo{series}{CHI EA '18})}. \bibinfo{publisher}{Association for
  Computing Machinery}, \bibinfo{address}{New York, NY, USA},
  \bibinfo{pages}{1–6}.
\newblock
\showISBNx{9781450356213}
\urldef\tempurl%
\url{https://doi.org/10.1145/3170427.3188575}
\showDOI{\tempurl}


\end{thebibliography}

\appendix
\presec
\section{Demographics of Recruited Participants}~\label{appenx:demo}\postsec
As a supplementary to our participant recruitment (see \S\ref{sec::methods::participants}), this section provides additional details on the demographic data of the recruited participants.
Table~\ref{tab::patient_data} demonstrates the demographics of $16$ recruited aging patients.
Although older adults usually have one or more chronic diseases due to the nature of the aging process, we only show self-reported significant comorbidities by system in Table~\ref{tab::patient_data}.
We decided to include this information to better contextualize participants self-assessment of QoL, specifically when it comes to {\it Prescription Management and Health Information}.
Table~\ref{tab::clinician_data} shows the demographics of $2$ geriatricians and $3$ nurses. 
Notably, all $5$ providers are affiliated with UC San Diego Health and all $16$ older adults live in San Diego and receive health care provided by the outpatient geriatric primary care clinic at UC San Diego Health.
As per required by IRB, we de-identified participants with their Personal Identifiable Information (PII) and Personal Health Information (PHI).

\begin{table}[h]
    \centering
    \includegraphics[width=0.48\textwidth]{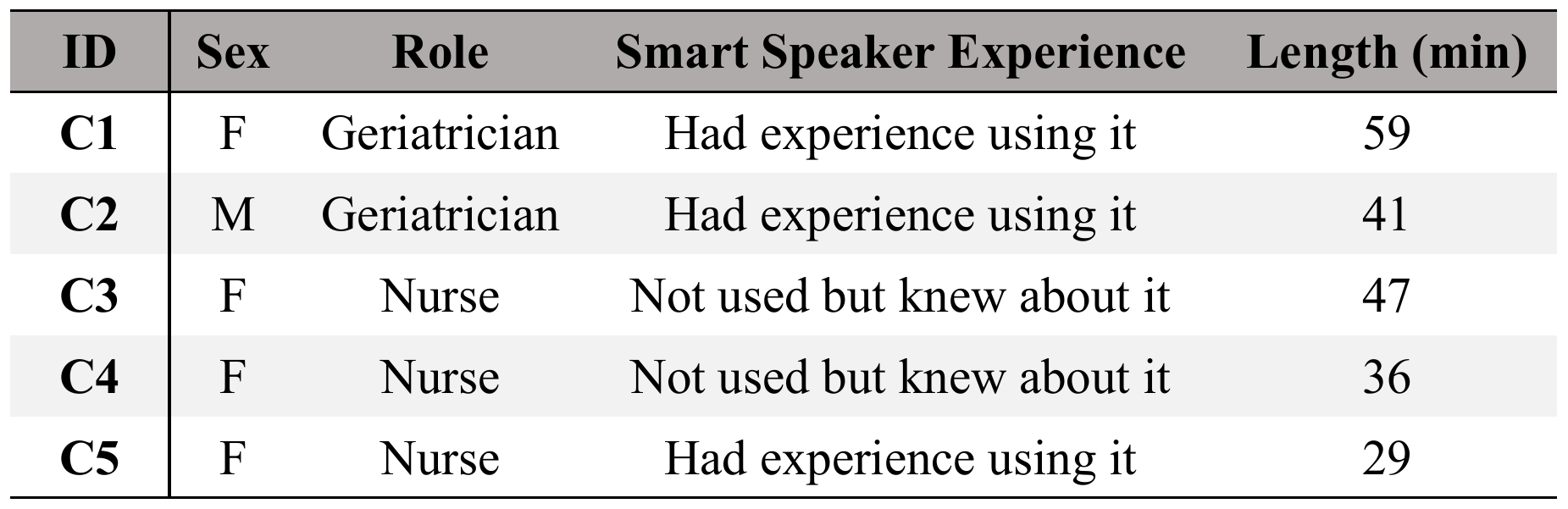}
    \Description{Demographics data of the recruited providers.}
    \captionof{table}{Demographics of $5$ recruited healthcare professionals.}
    \label{tab::clinician_data}
\end{table}

\begin{table*}[t]
    \centering
    \includegraphics[width=\textwidth]{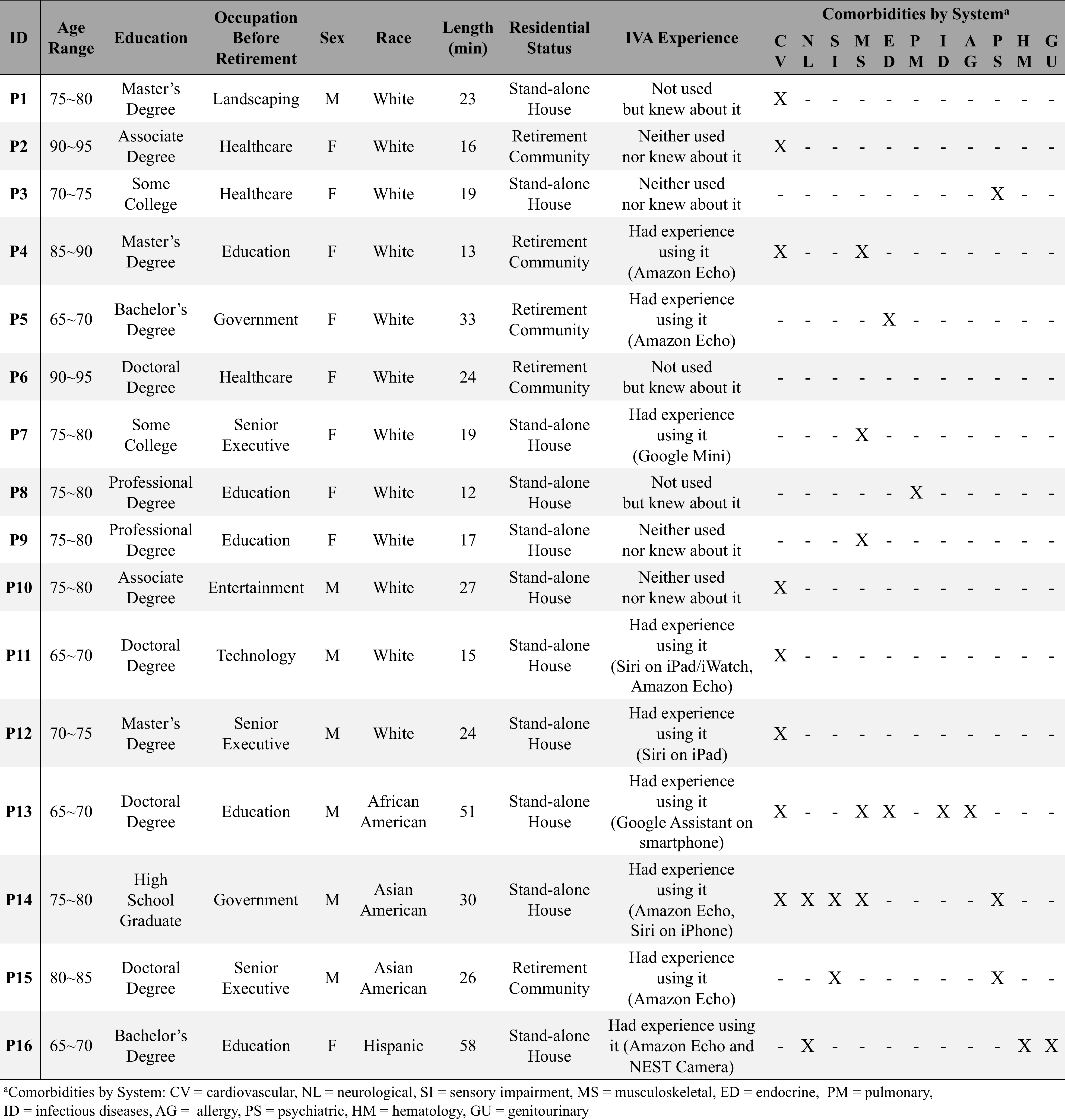}
    \Description{Demographics data of $16$ recruited older adults. The information include id, age range, education, occupation before retirement, sex, race, length of interview in min, residential status, previous IVA experience, and comorbidities by system.}
    \captionof{table}{Demographics of recruited older adults (age, $\mu = 75.56$~min, $\sigma = 6.97$~min).}
    \label{tab::patient_data}
\end{table*}

\end{document}